\documentclass[journal]{IEEEtran}
\usepackage{amsmath,graphicx,balance}
\usepackage{mathtools}
\usepackage{multirow}
\usepackage{subcaption}
\usepackage[noadjust]{cite}
\usepackage{xcolor,color}
\def\etal{et al.~}
\DeclareMathOperator*{\argmin}{arg\,min} 
\DeclareMathOperator*{\argmax}{arg\,max} 

\ifCLASSINFOpdf
\else
\fi
\begin{document}
\title{Domain Adversarial for Acoustic Emotion Recognition}

\author{Mohammed~Abdelwahab,~\IEEEmembership{Student Member,~IEEE,}
        and~Carlos~Busso,~\IEEEmembership{Senior~Member,~IEEE}% <-this % stops a space
\thanks{M. Abdelwahab and C. Busso are with the Department
of Electrical and Computer Engineering, The University of Texas at Dallas, Richardson
TX, 75013 USA e-mail: Mohammed.Abdel-Wahab@utdallas.edu; busso@utdallas.edu}% <-this % stops a space
\thanks{Manuscript received March 26, 2018; revised xxx xx, 2018.}}


\markboth{IEEE Transactions on Audio, Speech and Language Processing,~Vol.~xx, No.~xx, March~2018}%
{Abdelwahab and Busso: Domain Adversarial for Acoustic Emotion Recognition}



\maketitle

\begin{abstract}
The performance of speech emotion recognition is affected by the differences in data distributions between train (source domain) and test (target domain) sets used to build and evaluate the models. This is a common problem, as multiple studies have shown that the performance of emotional classifiers drop when they are exposed to data that does not match the distribution used to build the emotion classifiers. The difference in data distributions becomes very clear when the training and testing data come from different domains, causing a large performance gap between validation and testing performance. Due to the high cost of annotating new data and the abundance of unlabeled data, it is crucial to extract as much useful information as possible from the available unlabeled data. This study looks into the use of adversarial multitask training to extract a common representation between train and test domains. The primary task is to predict emotional attribute-based descriptors for arousal, valence, or dominance. The secondary task is to learn a common representation where the train and test domains cannot be distinguished. By using a gradient reversal layer, the gradients coming from the domain classifier are used to bring the source and target domain representations closer. We show that exploiting unlabeled data consistently leads to better emotion recognition performance across all emotional dimensions. We visualize the effect of adversarial training on the feature representation across the proposed deep learning architecture. The analysis shows that the data representations for the train and test domains converge as the data is passed to deeper layers of the network. We also evaluate the difference in performance when we use a shallow neural network versus a \emph{deep neural network} (DNN) and the effect of the number of shared layers used by the task  and domain classifiers.
\end{abstract}


% \fi
\IEEEpeerreviewmaketitle

\section{Introduction}
\IEEEPARstart{I}{n} many practical applications for speech emotion recognition systems, the testing data (target domain) is different from the labeled data used to train the models (source domain). The mismatch in data distribution leads to a performance degradation of the trained models \cite{Busso_2013,Kim_2014,Ververidis_2004,Vogt_2005,Parthasarathy_2017_3}. Therefore, it is vital to develop more robust systems that are more resilient to changes in train and test conditions \cite{Abdelwahab_2017, Abdelwahab_2017_2, Zong_2016, Abdelwahab_2015}. One approach to ensure that models perform well on the target domain is to use training data drawn from the same distribution. However, this approach can be expensive, since it requires enough data with emotional labels to build models specific to a new target domain. A more practical approach is to use available labeled data from similar domains along with unlabeled data from the target domain, creating models that generalize well to the new testing conditions without the need to annotate extra data with emotional labels. This study proposes an elegant solution for the problem of mismatch in train-test data distributions based on domain adversary training. 

\begin{figure}
\includegraphics[width=\columnwidth]{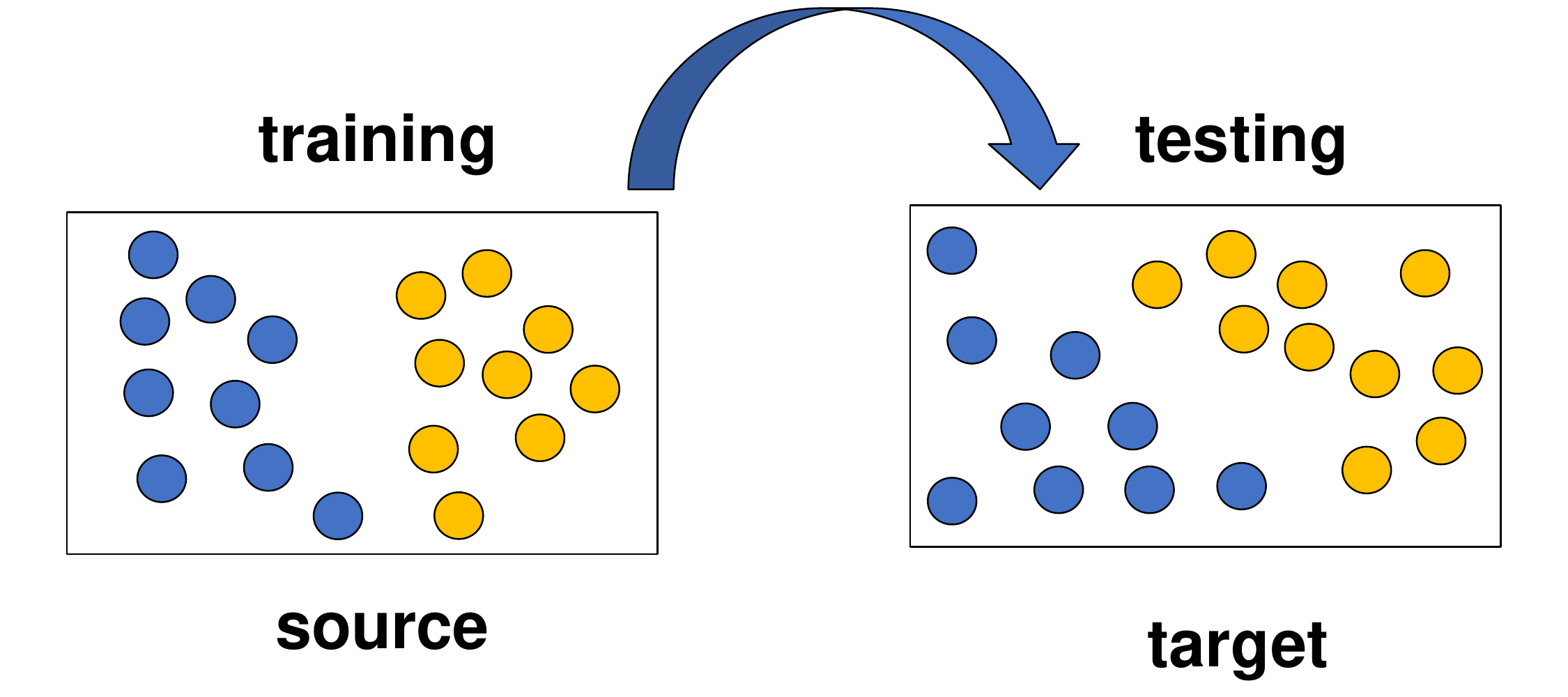}
\caption{Formulation of the problem of data mismatch between train (source domain) and test (target domain) datasets. Our solution uses unlabeled data from the target domain to reduce the differences between domains.}
\label{fig:prb}
\end{figure}

We formulate the machine-learning problem as follows. We have a \emph{source} domain(s) with annotated emotional data, which is used to train the models, and a large \emph{target} domain with unlabeled data (see Fig. \ref{fig:prb}). The testing data come from the target domain. Due to the prohibitive cost of annotating new data every time we change the target domain and the abundance of unlabeled data, we aim to use the unlabeled target data to extract useful information, reducing the differences between source and target domains. The envisioned system generalizes better and is more robust by maximizing the performance using a shared data representation for source and target domains. The key principle in our approach is to find a consistent feature representation for the source and target domains. Common approaches to address this problem include feature transformation methods, where the representation of the source data is mapped into a representation that resembles the features in the target domain \cite{Rahman_2012}, and finding a common representation between the domains, such that the features are invariant across domains \cite{Deng_2014_2, Zhang_2017, Glorot_2011}. The common domain-invariant features do not necessarily contain useful information about the main task. Therefore, it is vital to constrain the learned common feature representation to ensure that it is discriminative with respect to the main task. 

This paper explores the idea of finding a common representation between the source and target domains, such that data from the domains become indistinguishable while maintaining the critical information used in emotion recognition. This work is inspired by the work of Ganin \etal \cite{Ganin_2016}, which proposed training an adversarial multitask network. The approach searches the feature space for a representation that accurately describes the data from either domain, containing the relevant information to accurately classify the main task, in our study, the prediction of emotional attributes (arousal, valence, and dominance). The discriminative and domain-invariant features are learned by aligning the data distributions from the domains through back-propagation. This approach allows our framework to use unlabeled target data to learn a flexible representation.

This paper shows that adversarial training using unlabeled training data benefits emotion recognition. By using the abundant unlabeled target data, we gain on average 27.3\% relative improvement in \emph{concordance correlation coefficient} (CCC) compared to just training with the source data. We evaluate the effect of adversarial training by visualizing the similarity of the data representation learned by the network from both domains. The visualization shows that adversarial training aligns the data distributions as expected, reducing the gap between source and target domains. The study also shows the effect of the number of shared layers between the domain classifier and the emotion predictor on the performance of the system. The size of the source domain is an important factor that determines the optimal number of shared layers in the network. This novel framework for emotion recognition provides an appealing approach to effectively leverage unlabeled data from a new domain, generalizing the models and improving the classification performance. 

This paper is organized as follows. Section \ref{sec:relatedWork} discusses previous work on speech emotion recognition, with emphasis on frameworks that aim to reduce the mismatch between train and test datasets. Section \ref{sec:prop} presents our proposed model, providing the motivation and the details of the proposed framework. Section \ref{sec:proposed} presents the experiment evaluation including the databases, network structure and acoustic features. Section \ref{sec:print} presents the experimental results and the analysis of the main findings. Section \ref{sec:conc} finalizes the study with conclusions and future research directions.

\section{Related Work}
\label{sec:relatedWork}
%\subsection{Emotion Recognition}

The key challenge in speech emotion recognition is to build classifiers that perform well under various conditions. The cross-corpora evaluation in Shami and Verhels \cite{Shami_2007} demonstrated the drop in classification performance observed when training on one emotional corpus and testing on another. Other studies have shown similar results \cite{Austermann_2005, Schuller_2010_2, vidrascu_2008, Parthasarathy_2017_3}. Several approaches have been proposed to solve this problem. Shami and Verhels \cite{Shami_2007} proposed to include more variability in the training data by merging emotional databases. They demonstrated that it is possible to achieve classification performance comparable to within-corpus results. More recently, Chang and Scherer \cite{Chang_2017} showed that data from other domains can improve the within-corpus performance of a neural network. They employed \emph{deep convolutional generative adversarial network} (DCGAN) to extract and learn useful feature representation from unlabeled data from a different domain. This led to better generalization compared to a model that did not make use of unlabeled data. 
%{\color{red}[How close is this paper to our paper?]} {\color{blue} It is more similar to the current work however their work isn't necessarily cross-domain as the train and test corpus is IEMOCAP LOSO they also use AMI for the generator}

The main approach to attenuate the mismatch between train and test conditions is to minimize the differences in the feature space between both domains. Zhang \etal \cite{Zhang_2011_2} showed that by separately normalizing the features of each corpus, it is possible to minimize cross-corpus variability. Hassan \etal \cite{Hassan_2013} increased the weight of the train data that matches the test data distribution by using \emph{kernel mean matching} (KMM), \emph{Kullback-Leibler importance estimation procedure} (KLEIP), and \emph{unconstrained least-squares importance fitting} (uLSIF). Zong \etal \cite{Zong_2016} used \emph{least square regression} (LSR) to mitigate the projected mean and covariance differences between source data and unlabeled target samples while learning the regression coefficient matrix. Because the learned coefficient matrix depends on the samples selected from the target domain, multiple matrices were estimated and used to test new samples. Each matrix was used to predict an emotional label for a test sample, combining the results with the majority vote rule.

Studies have also explored mapping both train and test domains to a common space, where the feature representation is more robust to the variations between the domains. Deng \etal \cite{Deng_2014_2} used auto-encoders to find a common feature representation across the domains. They trained an auto-encoder such that it minimized the reconstruction error on both domains. Building upon this work, Mao \etal \cite{Mao_2016} proposed to learn a shared feature representation across domains by constraining their model to share the class priors across domains. Sagha \etal \cite{Sagha_2016}, also motivated by the work of Deng \etal \cite{Deng_2014_2}, used \emph{principal component analysis} (PCA) along with \emph{kernel canonical correlation analysis} (KCCA) to find views with the highest correlation between the source and target corpora. First, they used PCA to represent the feature space of the source and target data. Then, the features for the source and target domains were projected using the PCA in both domains. Finally, they used KCCA to select the top $N$ dimensions that maximized the correlation between the views. Inspired by universum learning where unlabeled data is used to regularize the training process for \emph{support vector machine} (SVM), Deng \etal \cite{Deng_2017} proposed adding an universum loss to the reconstruction loss of an auto-encoder. The added loss function was defined as the addition of the $L_2$-margin loss and the $\epsilon$-insensitive loss, making use of both labeled and unlabeled data. This approach aimed to learn auto-encoder classifier has low reconstruction and classification errors on both domains.

Song \etal \cite{Song_2016} proposed a couple of methods based on \emph{non-negative matrix factorization} (NMF) that utilized data from both train and test domains. The proposed methods aimed to represent a matrix formed by data from both domains as two non-negative matrices whose product is an approximation of the original matrix. The factorization was regularized by \emph{maximum mean discrepancy} (MMD) to ensure that the differences in the feature distributions of the two corpora were minimized. The proposed methods aim to learn a robust low dimensional feature representation using either unlabeled data or labels as hard constraints on the problem. Abdelwahab and Busso \cite{Abdelwahab_2017} proposed creating an ensemble of classifiers, where each classifier focuses on a different feature space (each classifier maximized the performance for a given emotion category). The features were selected over the labeled data from the target domain obtained with active learning. This semisupervised approach learned discriminant features for the target domain, increasing the robustness against shift in the data distributions between domains.

Instead of finding a common representation between domains, Deng \etal \cite{Deng_2013} trained a sparse auto-encoder on the target data and used it to reconstruct the source data. This approach used feature transformation in a way that exploits the underlying structure in emotional speech learned from the target data. Deng \etal \cite{Deng_2014} used two denoising auto-encoders. The first auto-encoder was trained on the target data and the second auto-encoder was trained on the source data, but it was constrained to be close to the first auto-encoder. The second auto-encoder was then used to reconstruct the data from both source and target domains. 

Our proposed approach takes an innovative formulation with respect to previous work relying on \emph{domain adversarial neural network} (DANN) \cite{Ganin_2016}. While Shinohara \cite{shinohara_2016} showed that the use of DANN can increase the robustness of a \emph{automatic speech recognition} (ASR) system against certain types of noise, this framework has not been used for speech emotion recognition, which is an important contribution as this framework can reduce the mismatch between train and test sets in a principled way. DANN relies on adversarial training for domain adaptation to learn a flexible representation during the training of the emotion classifier. As the training data changes, both the emotion and domain classifiers readjust their weights to find the new representation that satisfies all conditions. The domain classifier can be considered as a regularizer that prevents the main classifier from over-fitting to the source domain. The final learned representation performs well on the target domain without sacrificing the performance on the source domain. 

\section{Proposed Approach}
\label{sec:prop}
\subsection{Motivation}
\label{ssec:motiv}

We present an unsupervised approach to reduce the mismatch between source and target domains by creating a discriminative feature representation that leverages unlabeled data from the target domain. 

We aim to learn a common representation between the source and target domains, where samples from both domains are indistinguishable to a domain classifier. This approach is useful because all the knowledge learned while training the classifier on the source domain is directly applicable to the target domain data. We learn the representation by using a gradient reversal layer where the gradient produced by the domain classifier is multiplied by a negative value when it is propagated back to the shared layers. Changing the sign of the gradient causes the feature representation of the samples from both domains to converge, reducing the gap between domains. Ideally, the performance of the domain classifiers should be at random level where both domains ``look'' the same. When such a representation is learned, the data distributions of both domains are aligned. This approach leads to a large performance improvement in the target domain, as demonstrated by our experimental evaluation (see Section \ref{sec:print}). A key feature of this framework is that it is unsupervised, since it does not require labeled data from the target domain. However, this framework would continue to be useful when labeled data from the target domain is available, working as a semi-supervised approach.

\subsection{Domain Adversarial Neural Network for Emotion Recognition}
\label{ssec:dann}

\begin{figure}
\includegraphics[width=0.9\columnwidth]{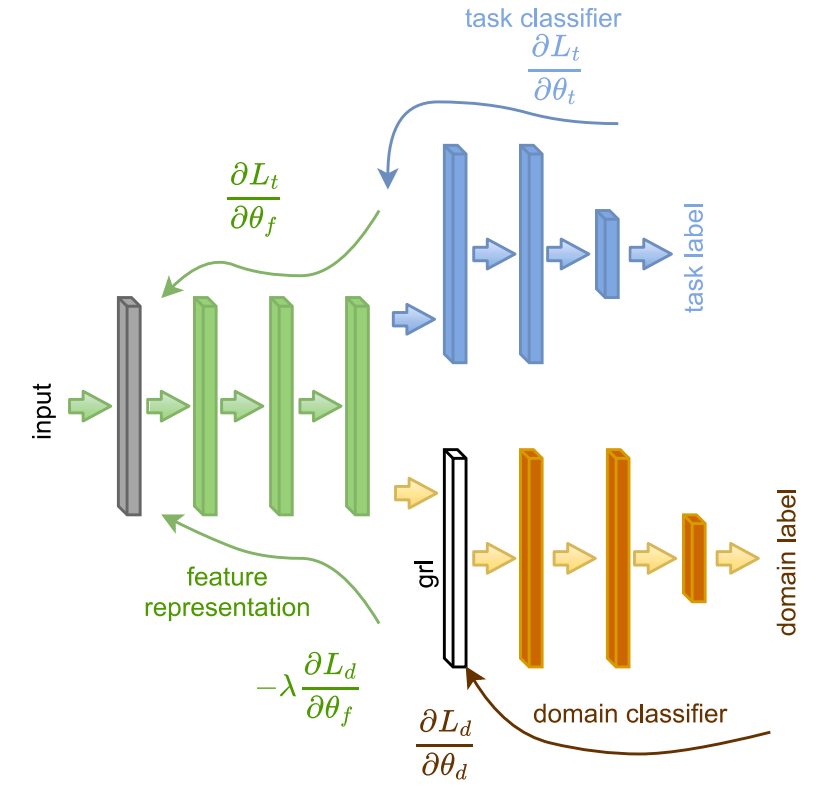}
\caption{Architecture of the \emph{domain adversarial neural network} (DANN) proposed for emotion recognition. The network has three parts: a feature representation common to both tasks, a task classifier layer, and a domain classifier layer.}
\label{fig:dann}
\end{figure}

Ganin \etal \cite{Ganin_2016}, inspired by the recent work on \emph{generative adversarial networks} (GAN) \cite{Goodfellow_2014}, proposed the \emph{domain adversarial neural network} (DANN). The network is trained using labeled data from the source domain and unlabeled data from the target domain. The network learns two classifiers: the main classification task, and the domain classifier, which determines whether the input sample is from the source or target domains. Both classifiers share the first few layers that determine the representation of the data used for classification. The approach introduced a gradient reversal layer between the domain classifier and the feature representation layers. The layer passes the data during forward propagation and inverts the sign of the gradient during backward propagation. The network attempts to minimize the task classification error and find a representation that maximizes the error of the domain classifier. By considering these two goals, the model ensures a discriminative representation for the main recognition task that makes the samples from either domain indistinguishable.

Figure \ref{fig:dann} shows an example structure of the DANN network. The network is fed labeled source data and unlabeled target data in equal proportions. In our formulation, we propose to predict emotional attribute descriptors as the primary goal. We train the primary recognition task with the source data, for which we have emotional labels. For the domain classifier, we train the classifier with data from the source and target domains. Notice that the domain classifier does not require emotional label, so we can rely on unlabeled data from the target domain. The classifiers are trained in parallel. The network's objective is defined as:

\begin{equation}
\begin{split}
\MoveEqLeft
E(\theta_f , \theta_y, \theta_d) = \frac{1}{n} \sum_{i=1}^{n} L_y^i(\theta_f , \theta_y)\\
&- \lambda \Big(\frac{1}{n}\sum_{i=1}^{n}L_d^i(\theta_f,\theta_d) + \frac{1}{m}\sum_{i=1}^{m}L_d^i(\theta_f,\theta_d)\Big)
\end{split}
\label{eq:E}
\end{equation}

\noindent
where $\theta_f$ represents the parameters of the shared layers providing the regularized feature representation, $\theta_y$ represents the parameters of the layers associated with the main prediction task, and $\theta_d$ represents the parameters of the layers of the domain classifier ($n$ is the number of labeled training samples, $m$ is the number of unlabeled training samples). The optimization process consists of two loss functions. $ L_y^i$ is the prediction loss for the main task, and $ L_d^i$ is the domain classification loss. The prediction loss and the domain classification loss compete against each other in an adversarial manner. The parameter $\lambda$ is a regularization multiplier that controls the tradeoff between the losses. This is a minimax problem. It attempts to find a saddle point parametrized by $\hat{\theta}_f,\hat{\theta}_y,\hat{\theta}_d$, 

\begin{equation}
(\hat{\theta}_f,\hat{\theta}_y) = \argmin_{\theta_f,\theta_y} E(\theta_f,\theta_y,\hat{\theta}_d)
\end{equation}
\begin{equation}
\hat{\theta}_d = \argmax_{\theta_d} E(\hat{\theta}_f,\hat{\theta}_y, \theta_d)
\end{equation}

At the saddle point, the classification loss on the source domain is minimized and the domain classification loss is maximized. The maximization is achieved by introducing a gradient reversal layer that changes the sign of the gradient going from the domain classification layers to the feature representation layers (see white layer in Fig. \ref{fig:dann}). The updates taken on the feature representation parameters are in opposite direction to the gradient. With this approach, stochastic gradient descent tries to make the features similar across domains, so what is learned from the source domain remains effective for the target domain without loss in performance.

The simple concept of choosing a representation that confuses a competent domain classifier leads to models that perform better in the target domain without impacting the performance in the source domain. This approach is particularly useful for emotion recognition, as most annotated corpora come from studio settings that greatly differs from real world testing conditions. This unsupervised approach uses available unlabeled data to align the distributions of both domains. The aligned distributions lead to a common representation, causing the domain classifier's performance to drop to random chance levels. The common indistinguishable representation retains discriminative information learned during the training of the models with data from the source domain. We improve the classifier's performance on the target domain without having to collect new annotated data. This is an important contribution in this field, taking us one step closer toward robust speech emotion classifiers that generalize well in most testing conditions.

\section{Experimental Evaluation}
\label{sec:proposed}

We define the main task as a regression problem to estimate the emotional content conveyed in speech described by the emotional attributes arousal (calm versus activated), valence (negative versus positive), and dominance (weak versus strong). This section describes the databases (Section \ref{ssec:db}), the acoustic features (Section \ref{ssec:feat}) and the specific network structures (Section \ref{ssec:net}) used in the experimental evaluation.  

\subsection{Emotional Databases}
\label{ssec:db}

The experimental evaluation consider a multi-corpus setting with three databases. The source domain (test set) corresponds to two databases: the USC-IEMOCAP \cite{Busso_2008_5} and MSP-IMPROV \cite{Busso_2017} corpora. The target domain corresponds to the MSP-Podcast \cite{Lotfian_201x} database.

\subsubsection{The USC-IEMOCAP Corpus}
\label{sssec:iem}

The USC-IEMOCAP database is an audiovisual corpus recorded from ten actors during dyadic interactions \cite{Busso_2008_5}. It has approximately 12 hours of recordings with detailed motion capture information carefully synchronized with audio (this study only uses the audio). The goal of the data collection was to elicit natural emotions within a controlled setting. This goal was achieved with two elicitation frameworks: emotional scripts, and improvisation of hypothetical scenarios. These approaches allowed the actors to express spontaneous emotional behaviors driven by the context, as opposed to read speech displaying prototypical emotions \cite{Busso_2008}. Several dyadic interactions were recorded and manually segmented into turns. Each turn was annotated with emotional labels by at least two evaluators across emotional attributes (valence, arousal, dominance). Dimensional attributes take integer values that range from one to five. The dimensional attribute of an utterance is the average of the values given by the annotators.
We linearly map the scores between $-3$ and 3.

\subsubsection{The MSP-IMPROV Corpus}
\label{sssec:imp}

The MSP-IMPROV database is a multimodal emotional database recorded from actors interacting in dyadic sessions \cite{Busso_2017}. The recordings were carefully designed to promote natural emotional behaviors, while maintaining control over lexical and emotional contents. The corpus relied on a novel elicitation scheme, where two actors improvised scenarios that lead one of them to utter target sentences. For each of these target sentences, four emotional scenarios were created to contextualize the sentence to elicit happy, angry, sad and neutral reactions, respectively. The approach allows the actor to express emotions as dictated by the scenarios, avoiding prototypical reactions that are characteristic of other acted emotional corpus. Busso \etal \cite{Busso_2017} showed that the target sentences occurring within these improvised dyadic interactions were perceived more natural than read renditions of the same sentences. The MSP-IMPROV corpus includes not only the target sentences, but also other sentences during the improvisations that led one of the actors to utter the target sentence. It also includes the natural interactions between the actors during the breaks. 

The corpus consists of 8,438 turns of emotional sentences recorded from 12 actors (over 9 hours). The sessions were manually segmented into speaking turns, which were annotated with emotional labels using perceptual evaluations conducted with crowdsourcing \cite{Burmania_2016_2}. Each turn was annotated by at least five evaluators, who annotated the emotional content in terms of the  dimensional attributes arousal, valence, and dominance. Dimensional attributes take integer values that range from one to five. The consensus label assigned to each speech turn is the average value of the scores provided by the evaluators, which we linearly map between $-3$ and 3.

\subsubsection{The MSP-Podcast Corpus}
\label{sssec:pod}

The \emph{MSP-Podcast} corpus is an extensive collection of natural speech from multiple speakers appearing in Creative Commons licensed recordings downloaded from audio-sharing websites \cite{Lotfian_201x}. Some of the key aspects of the corpus are the different conditions in which the recordings are collected, large number of speakers, and a large variety of natural content from spontaneous conversations conveying emotional behaviors. The audio was preprocessed removing portions that contain noise, music or overlapped speech. The recordings were then segmented into speaking turns creating a big audio repository with sentences that are between 2.75 seconds and 11 seconds. Emotional models trained with existing databases are then used to retrieve speech turns with target emotional content. The candidate segments were annotated with emotional labels using an improved version of the crowdsourcing framework proposed by Burmania \etal \cite{Burmania_2016_2}. 

Each speech segment was annotated by at least five raters, who provided scores for the emotional dimensions arousal, valence and dominance using seven-likert scales. The consensus scores are the average scores assigned by the evaluators, which are shifted such that they are in the range between $-3$ and 3. The collection of this corpus is an ongoing effort. This study uses 14,227 labeled sentences. From this set, we use 4,283 labeled sentences coming from 50 speakers as our test set, which is consistently used across conditions. For the within corpus evaluation (i.e., training and testing in the same domain), we define a development set with 1,860 sentences from 10 speakers, and a train set with the remainder of the corpus (8,084 sentences). This study also uses 73,209 unlabeled sentences from the audio repository of segmented speech turns. The unlabeled segments are used to train the domain classifier in the DANN approach.

\subsection{Acoustic Features}
\label{ssec:feat}

The acoustic features correspond to the set proposed for the INTERSPEECH 2013 Computational Paralinguistics Challenge (ComParE) \cite{Schuller_2013}. This feature set includes 6,373 acoustic features extracted at the sentence level (one feature vector per sentence). First, it extracts 65 frame-by-frame \emph{low-level descriptors} (LLDs) which includes various acoustic characteristics such as \emph{Mel-frequency cepstral coefficients} (MFCCs), fundamental frequency, and energy. The externalization of emotion is conveyed through different aspects of speech production so including these LLDs is important to capture emotional cues. After estimating LLDs, a list of functions are estimated for each LLD, which are referred to as \emph{high-level descriptors} (HLD) features. These HLDs include standard deviation, minimum, maximum, and ranges. The acoustic features are extracted using OpenSMILE \cite{Eyben_2010_2}.

We separately normalize the features from each domain (i.e., corpus) to have zero mean and a unit standard deviation. The mean and the variance of the data is calculated considering only the values of the features within the 5\% and 95\% quantiles to avoid outliers skewing the values. After normalization, we ignore any value greater than 10 times its standard deviation by setting their values to zero.

\subsection{Network Structure}
\label{ssec:net}
As discussed in Section \ref{ssec:dann}, the DANN approach has three main components: the domain classifier layers, task classifier layers, and feature representation layers. The \emph{domain classifier layers} are implemented with two layers across all the experimental evaluation. The \emph{task classifier layers} are also implemented with two layers, except for the shallow network described below. The number of layers in the \emph{feature representation layers} is a parameter of the network that is set on the development set. We consider different number of shared layers, evaluating the performance of the system with one, two, three or four layers. 

We also study whether a simple shallow network can achieve similar performance compared to the deep network. In the shallow network, the \emph{task classifier layer} and the \emph{feature representation layer} are implemented with one layer each. We implement the \emph{domain classifier layer} with two layers.

% {\color{red} [Check later]
% We also study whether a simple shallow network can achieve similar performance compared to the deep network. In the shallow network, we have one shared layer and task classification layer. Since the performance of the adversarial training depends on the capacity of the domain classifier \cite{Ganin_2016}, where deeper networks have higher capacity, yielding better results compared to shallow networks with low capacity. We keep the domain classifier to two layers.}

%we consider two network structures: a shallow structure, consisting of one layer, and a deep structure consisting of four layers. The performance of the adversarial training depends on the capacity of the domain classifier \cite{Ganin_2016}. Deeper networks have higher capacity, yielding better results compared to shallow networks with low capacity. We fix the domain classifier to two layers across structures.
%{\color{red} [How many layers for the main task? -- blue layers in fig 2?]} {\color{blue} The total number of layers from the input to the main task output is four, the number of layers for the blue part of the network varies depending on the number of layers in the shared network part}
\subsection{Baseline Systems}
\label{ssec:base}
We establish two baselines. The first baseline is a network trained and validated only on the source data. This condition creates a mismatch between the train and test conditions. The second baseline corresponds to within corpus evaluations, where the models are trained and tested with data from the target domain. This model assumes that training data from the target domain is available, so it corresponds to the ideal condition. The parameters of the networks are optimized using the development set. The baselines are implemented with similar architectures, serving as a fair comparison with the proposed method (e.g., number of layers, number of nodes). The key difference with the DANN models is the lack of the \emph{domain classification layers}, where the feature representation layers only consider the primary classification task.

%(1,860 sentences) using the training set (8,084 sentences)

\subsection{Implementation}
\label{ssec:Implementation}
We train the networks using Keras \cite{chollet_2015} with Tensorflow as back-end \cite{Abadi_2016}. We use batch normalization and dropout. The dropout rates are $p$=0.2 for the input layer and $p$=0.5 for the rest of the layers. We further regularize the models using max-norm on the weights of value four and a clip norm on the gradient of value ten. The loss function used for the main regression task is the \emph{mean square error} (MSE). The loss function for the domain classification task is the categorical cross-entropy. We use Adam as an optimizer with a learning rate of 5$e^{-4}$ \cite{Dozat_2015}. We train the models for 100 epochs with a batch size of 256 sentences. A parameter of the DANN model is $\lambda$ in Equation \ref{eq:E}, which controls the tradeoff between the task and domain classification losses. We follow a similar approach to the one proposed by Ganin \etal~\cite{Ganin_2016}, where $\lambda$ is initialized equal to zero for the first ten training epochs. Then, we slowly increase its value until reaching $\lambda=1$ by the end of the training. We train each model twenty times to reduce the effect of initialization on the performance of the classifiers. We report the average performance across the trails.

In adversarial training, we need unlabeled data to train the domain classifier. We randomly select samples from the unlabeled portion of the target domain to be fed to the domain classifier. The number of selected samples from the audio repository of unlabeled speech turns is equal to the number of samples in the source domain, keeping the balance in the training set of the domain classification task.

\section{Experimental Results}
\label{sec:print}

The performance results for the baseline models and the DANN models are reported in terms of the \emph{root mean square error} (RMSE), \emph{Pearson's correlation coefficient} (PR), and \emph{concordance correlation coefficient} (CCC) between the ground-truth labels and the estimated values. While we presented PR and RMSE, the analysis focus on CCC as the performance metric, which combines \emph{mean square error} (MSE) and PR. CCC is defined as:

\begin{equation}
\rho_c = \frac{2 \rho \sigma_x \sigma_y}{\sigma_{x}^2 + \sigma_{y}^2 + (\mu_x -\mu_y)^2} 
\end{equation}

\noindent
where $\rho$ is the Pearson's correlation coefficient, $\sigma_x$ and $\sigma_y$, and $\mu_x$ and $\mu_y$ are the standard deviations and means of the predicted score $x$ and the ground truth label $y$, respectively.

\subsection{Number of Layers For the Shared Feature Representation}
\label{ssec:depth}

We first study the effect of the number of shared layers between the domain classifier and the primary regression task on the DANN model's performance (e.g., \emph{feature representation layers} in Fig. \ref{fig:dann}). The domain and task classifier layers are implemented with two layers each. We vary the number of shared layers between the classifiers and observe how the changes in feature representation affect the regression performance. This evaluation is conducted exclusively on the validation set of the target domain (e.g., MSP-Podcast corpus). 

Table \ref{tbl:depth} shows the average RMSE, PR, and CCC for the models trained with one, two, three and four shared layers. For arousal, we consistently observe better performance (lower RMSE and higher PR and CCC) with one shared layer between the domain and task classifiers. For the CCC values, the differences are statistically significant for all the cases, except when the source domain is the MSP-IMPROV corpus and the DANN model is implemented with two layers (one-tailed t-test over the average across the twenty trails, asserting significance if $p$-value $< 0.05$). The performance degrades as more shared layers are added. For valence and dominance, the number of shared layers that provides the best performance varies from one corpus to another. In most cases, two or three shared layers provide the best performance. Based on these results on the validation set, we set the number of shared layers for the feature representation networks to one for arousal, two for valence and three for dominance. 

\begin{table}[thb]
\fontsize{7.5}{9}\selectfont
\centering
\caption{Performance of DANN framework implemented with different number of shared layers for the domain and task classifiers [\emph{Iem}: USC-IEMOCAP corpus, \emph{Imp}: MSP-IMPROV corpus, \emph{All}: All corpora combined]}
\label{tbl:depth}
\begin{tabular}{c|c|c@{\hspace{0.1cm}}c@{\hspace{0.1cm}}c|c@{\hspace{0.1cm}}c@{\hspace{0.1cm}}c|c@{\hspace{0.1cm}}c@{\hspace{0.1cm}}c}
\hline
\multicolumn{2}{c|}{} & \multicolumn{3}{c|}{Arousal} & \multicolumn{3}{c|}{Valence} & \multicolumn{3}{c}{Dominance} \\ \hline
src                   & n & RMSE & PR & CCC & RMSE & PR & CCC & RMSE & PR & CCC \\ \hline \hline
\multirow{4}{*}{Iem}     & 1      & 1.26  & .412  & .365  & 1.57  & .140  & .135  & 1.30  & .151  & .124 \\
                         & 2      & 1.29  & .379  & .332  & 1.64  & .152  & .144  & 1.27  & .160  & .135 \\
                         & 3      & 1.32  & .353  & .305  & 1.67  & .150   & .140  & 1.27  & .217  & .181 \\
                         & 4      & 1.31  & .347  & .296  & 1.67  & .147  & .135   & 1.27 & .192  & .161 \\ \hline
                         
\multirow{4}{*}{Imp}     & 1      & 1.62  & .497  & .284  & 1.58  & .171  & .137  & 0.80  & .485  & .448 \\
                         & 2      & 1.57  & .506  & .303  & 1.47  & .203  & .176  & 0.82  & .514  & .476 \\
                         & 3      & 1.71  & .399  & .218  & 1.48  & .202  & .173  & 0.87  & .478  & .403 \\
                         & 4      & 1.73  & .382  & .202  & 1.43  & .210  & .193  & 0.85  & .442  & .383 \\ \hline

\multirow{4}{*}{All}     & 1      & 1.37  & .432  & .341  & 1.56  & .181  & .165  & 1.01  & .395  & .356  \\
                         & 2      & 1.41  & .396  & .313  & 1.56  & .173  & .160  & 1.05  & .345  & .309  \\
                         & 3      & 1.46  & .369  & .286  & 1.58  & .183  & .170  & 1.06  & .339  & .308  \\
                         & 4      & 1.45  & .371  & .282  & 1.56  & .161  & .149  & 1.06  & .307  & .270  \\ \hline
\end{tabular}
\end{table}

\subsection{Regression Performance of the DANN Model}
\label{ssec:page}

This section presents the regression results achieved by the DANN model, and the baseline models. Table \ref{tab:1} lists the performance for the within-corpus evaluation where the models are trained and tested with data from the MSP-Podcast corpus (referred to as \emph{target} on the table), and the cross-corpus evaluations, where the models are trained with other corpora (referred to as \emph{src} on the table). These baseline results are compared with our proposed DANN method (referred to as \emph{dann} on the table). The results are tabulated in terms of emotional dimensions and networks structures. These values are the average of twenty trails to account for different initializations and set of unlabeled data used to train the DANN models. For the rows denoted ``All Databases'', we combine all the source domain together (IEMOCAP and MSP-IMPROV corpora), treating them as a single source domain. To determine statistical differences between the \emph{src} and \emph{dann} conditions, we use the one-tailed t-test over the twenty trails, asserting significance if $p$-value $< 0.05$. We highlight in bold when the difference between these conditions is statistically significant.

\begin{figure}
\centering
\includegraphics[width=0.9\columnwidth]{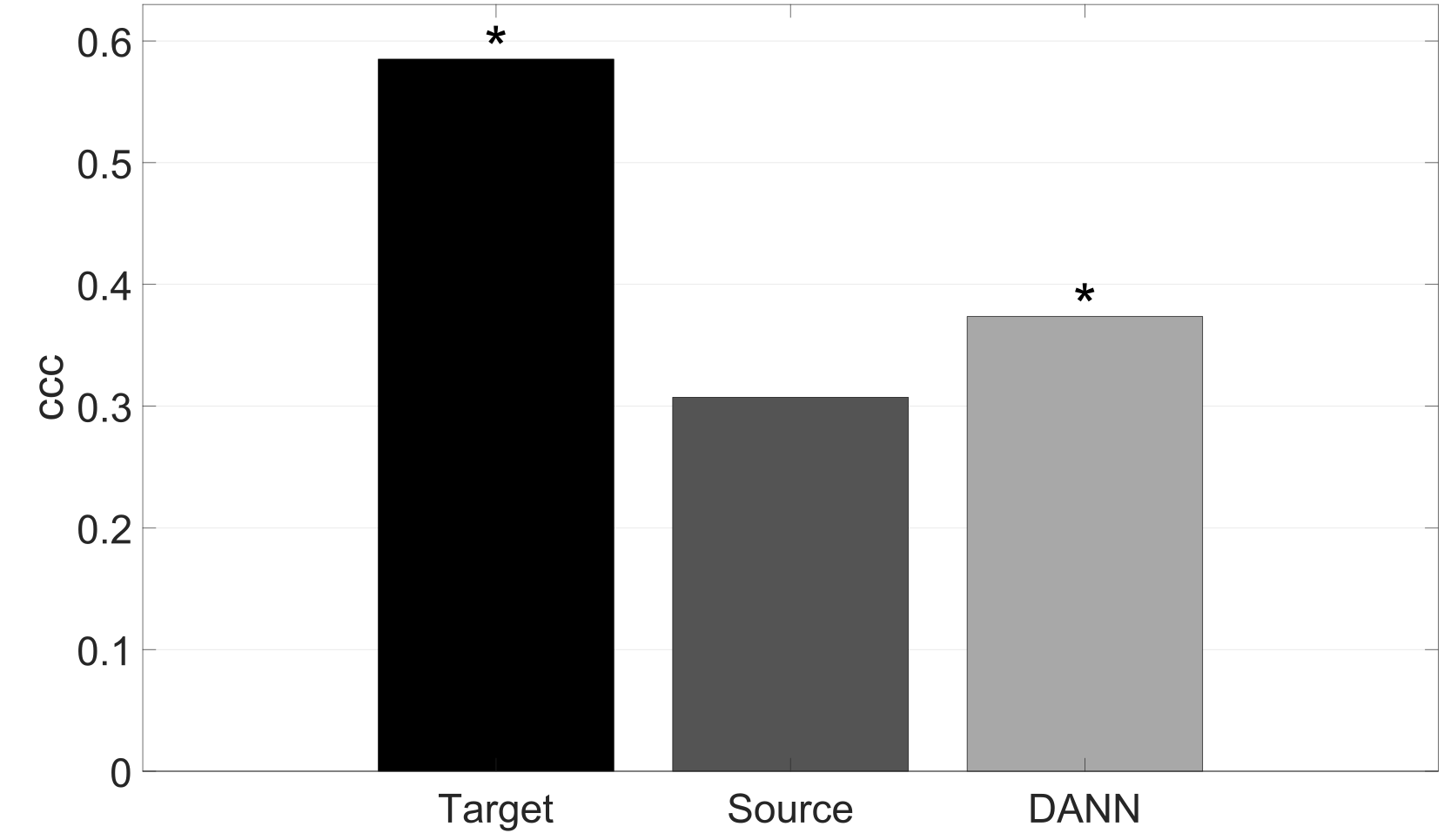}
\caption{Average concordance correlation coefficient across conditions. The DANN models provide significant improvements over the baseline model trained with the source domain.}
\label{fig:avg}
\end{figure}

\begin{figure}
\centering
\includegraphics[width=0.9\columnwidth]{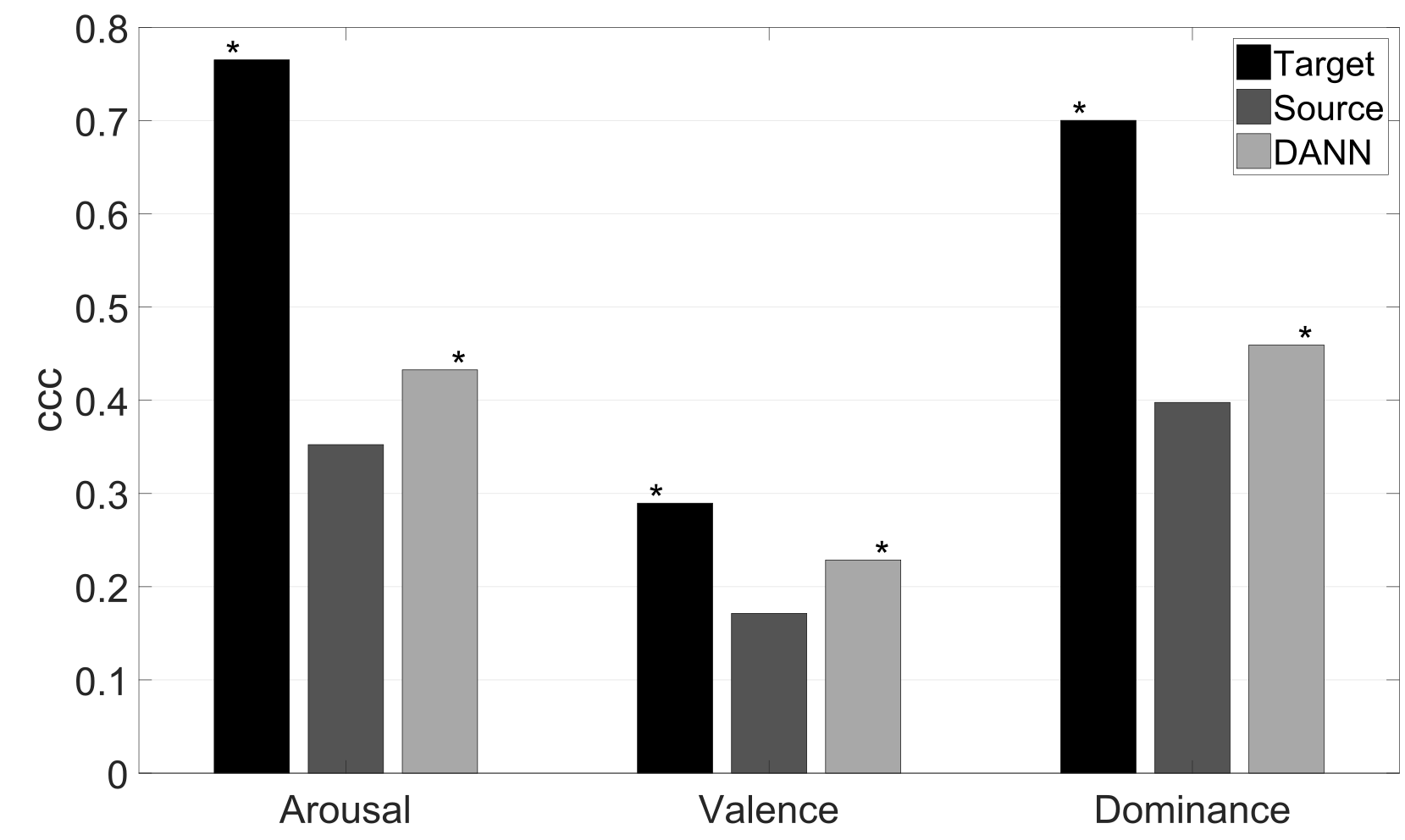}
\caption{Average concordance correlation coefficient across conditions for each  emotional dimension.}
\label{fig:avgdim}
\end{figure}

\begin{figure}
\centering
\includegraphics[width=0.9\columnwidth]{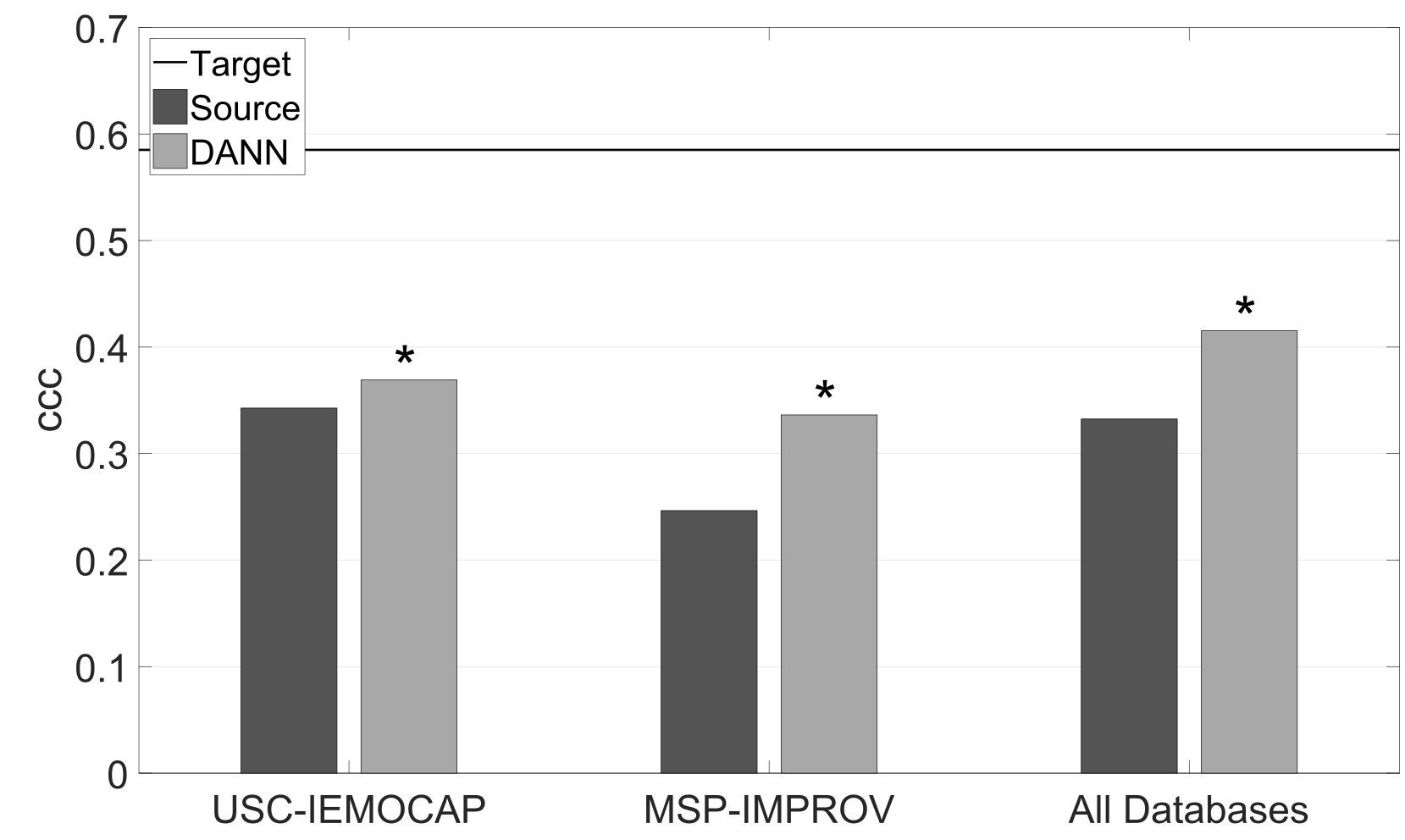}
\caption{Average concordance correlation coefficient across condition for each source domain. The solid line represents the within corpus performance.}
\label{fig:db}
\end{figure}

\begin{figure}
\centering
\includegraphics[width=0.9\columnwidth]{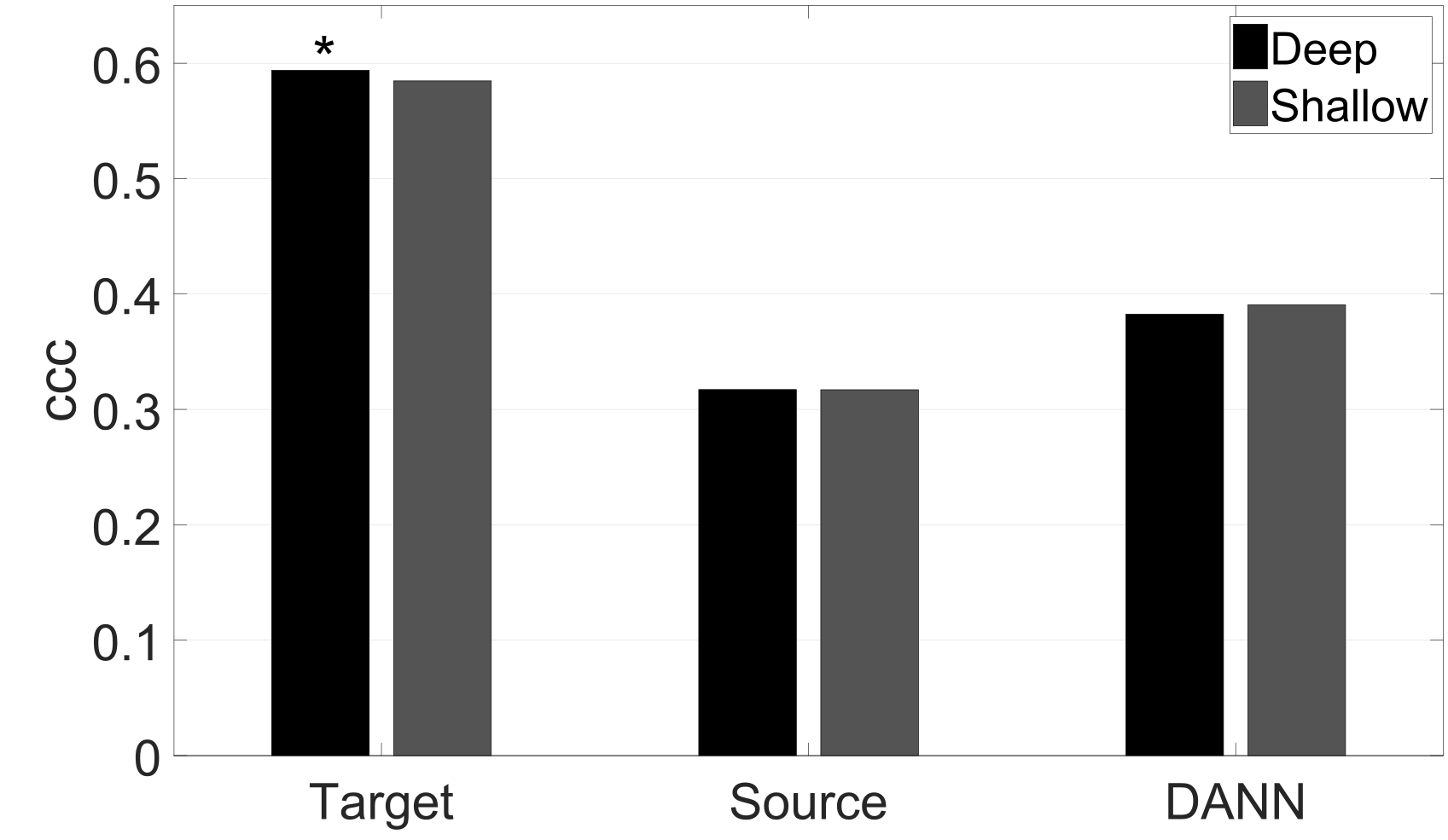}
\caption{Average concordance correlation coefficient across conditions for deep and shallow structures.}
\vspace{-0.2cm}
\label{fig:dpsh}
\end{figure}

To visualize the results in Table \ref{tab:1}, we create figures showing the average performance under different conditions (Figs. \ref{fig:avg}-\ref{fig:dpsh}). We use statistical significance tests to compare different conditions (values obtained from Table \ref{tab:1}). We test the hypothesis that population means for matched conditions are different using one-tailed z-test. We assert significance if $p$-value $< 0.05$. We use an asterisk in the figures to indicate if there is a statistically significant difference relative to the baseline model trained with the source domain.
% statistical differences here are for the figures not the table

\begin{table}[thb]
\centering
\fontsize{7.5}{9}\selectfont
\caption{Performance of the proposed DANN approach across different structures and emotional dimensions.}
\label{tab:1}
\begin{tabular}{l@{\hspace{0.1cm}}|@{\hspace{0.1cm}}c@{\hspace{0.1cm}}|c@{\hspace{0.1cm}}|c|c|c|@{\hspace{0.1cm}}c|c}
\hline 
\multicolumn{2}{c|}{Training Data} &  \multicolumn{6}{c}{Arousal} \\ 
\hline 
 \multirow{2}{*}{Source} & \multirow{2}{*}{Approach}
 & \multicolumn{3}{c|}{deep} & \multicolumn{3}{c}{shallow} \\ 
\cline{3-8} 
& & RMSE & PR & CCC &  RMSE & PR & CCC \\ 
\hline
\hline
MSP-Podcast & target & 0.67 & .786 & .776 & 0.66 & .787 & .767 \\
\hline	
\multirow{2}{*}{USC-IEMOCAP}
 & src & \textbf{1.01} & \textbf{.515} & .452 & \textbf{1.00} & .510 & .434 \\
 & dann & 1.10 & .503 & \textbf{.489} & 1.07 & \textbf{.520} & \textbf{.503} \\
 \hline
\multirow{2}{*}{MSP-IMPROV}
 & src & 1.57 & .551 & .267 & 1.53 & .555 & .263 \\
 & dann &\textbf{1.48} & \textbf{.607} & \textbf{.381} & \textbf{1.46} & \textbf{.614} & \textbf{.381} \\ 
 \hline
%\multirow{2}{*}{SEMAINE}
% & src & 1.08 & .508 & .295 & 1.11 & .489 & .230 \\
% & dann &1.05 & .540 & \textbf{.398} & 1.12 & .481 & .267\\ 
% \hline
\multirow{2}{*}{All Databases}
 & src & 1.20 & .496 & .386 & 1.18 & .506 & .378\\
 & dann &\textbf{1.18} & \textbf{.555} & \textbf{.499} & 1.18 & \textbf{.551} & \textbf{.486}\\ 
\hline
\multicolumn{8}{c}{}\\
\multicolumn{8}{c}{}\\
\hline
\multicolumn{2}{c|}{Training Data} &  \multicolumn{6}{c}{Valence} \\ 
\hline 
 \multirow{2}{*}{Source} & \multirow{2}{*}{Approach}
 & \multicolumn{3}{c|}{deep} & \multicolumn{3}{c}{shallow} \\ 
\cline{3-8} 
& & RMSE & PR & CCC &  RMSE & PR & CCC \\ 
\hline
\hline
MSP-Podcast & target & 1.08 & .327 & .294 & 1.03 & .344 & .283\\
\hline
\multirow{2}{*}{USC-IEMOCAP}
 & src & \textbf{1.30} & .202 & .198 & \textbf{1.19} & .227 & .207 \\
& dann &1.44 & \textbf{.218} & \textbf{.215} & 1.34 & \textbf{.267} & \textbf{.255}\\
 \hline
 \multirow{2}{*}{MSP-IMPROV}
 & src & \textbf{1.42} & .142 & .122 & \textbf{1.29} & .154 & .127 \\
 & dann &1.43 & \textbf{.163} & \textbf{.161} & 1.37 & \textbf{.180} & \textbf{.178}  \\
 \hline
\multirow{2}{*}{All Databases}
 & src & \textbf{1.33} & .214 & .201 & \textbf{1.16} & .245 & .228\\
 & dann &1.39 & \textbf{.299} & \textbf{.294} & 1.33 & \textbf{.272} & \textbf{.267} \\ 
\hline
\multicolumn{8}{c}{}\\
\multicolumn{8}{c}{}\\

\hline
\multicolumn{2}{c|}{Training Data} &  \multicolumn{6}{c}{Dominance} \\ 
\hline 
 \multirow{2}{*}{Source} & \multirow{2}{*}{Approach}
 & \multicolumn{3}{c|}{deep} & \multicolumn{3}{c}{shallow} \\ 
\cline{3-8} 
& & RMSE & PR & CCC &  RMSE & PR & CCC \\ 
\hline
\hline
MSP-Podcast & target & 0.57 & .718 & .697 & 0.57 & .723 & .704 \\
\hline
\multirow{2}{*}{USC-IEMOCAP}
& src & \textbf{0.95} & .437 & .393 & \textbf{0.87} & .461 & .426\\
& dann &1.10 & .437 & .401 & 1.03 & \textbf{.497} & \textbf{.457}\\
 \hline
 \multirow{2}{*}{MSP-IMPROV}
 & src & \textbf{0.86} & .563 & .368 & 0.87 & .520 & .353\\
 & dann &0.89 & .565 & \textbf{.456} & 0.88 & \textbf{.578} & \textbf{.472}\\
 \hline
\multirow{2}{*}{All Databases}
 & src & \textbf{0.85} & .481 & .418 & \textbf{0.81} & .493 & .437\\
 & dann &0.92 & \textbf{.526} & \textbf{.499} & 0.90 & \textbf{.550} & \textbf{.516}\\ 
 \hline 
\end{tabular}
\end{table}

Figure \ref{fig:avg} shows the average concordance correlation coefficient across emotional dimensions, training sources, trails, and networks structures (three emotional dimensions $\times$ twenty trails $\times$ three sources $\times$ five structures = 900 matched conditions).  The average performance for the within-corpus evaluation (target) is close to double the performance for the cross-corpus evaluations (source).  This result demonstrates the importance of minimizing the mismatch between train and test conditions in emotion recognition. The figure shows that the proposed DANN approach greatly improves the performance, achieving 6.6\% gain compared to the systems trained with the source domains. As highlighted by the asterisk, the improvement is statistically significant. The proposed approach reduces the gap between within-corpus and cross-corpus emotion recognition results, effectively using unlabeled data from the target domain.

Figure \ref{fig:avgdim} shows the average concordance correlation coefficient for each emotional dimension (twenty trails $\times$ three sources $\times$ five structures = 300 matched conditions). On average, the figure shows that models trained using DANN consistently outperform models trained with the source data. The asterisk denotes that the difference is statistically significant across all emotional dimensions. The relative improvements over the source models are 22.8\% for arousal, 33.4\% for valence and 15.5\% for dominance. In general, the values for CCC are lower for valence, validating findings from previous studies, which indicated that acoustic features are less discriminative for this emotional attribute \cite{Busso_2012}.

Figure \ref{fig:db} shows the average concordance correlation coefficient per source domain (three emotional dimensions $\times$ twenty trails $\times$ five structures = 300 matched conditions). The figure shows the within-corpus performance (target) with a solid horizontal line. The results consistently show significant improvements when using DANN. The relative improvements in performance over training with the source domain are 7.7\% for the USC-IEMOCAP corpus, 36.4\% for the MSP-IMPROV corpus, and 25\% when we combine all the corpora. Figure \ref{fig:db} also shows that, on average, combining all the sources into one domain improves the performance of the systems in recognizing emotions. Adding variability during the training of the models is important, as also demonstrated by Shami and Verhels \cite{Shami_2007}. DANN models also benefit from adding variability. By leveraging the added data representations, DANN models are able to find a common representation between the domains without sacrificing the critical features relevant for learning the primary task.

Figure \ref{fig:dpsh} compares the performance for deep and shallow networks (see Section \ref{ssec:net}). The figure summarizes the results for the within-corpus evaluations (target), cross-corpus evaluations (source) and with our proposed DANN model (three emotional dimensions $\times$ twenty trails $\times$ three sources = 180 matched conditions). For the target models, we observe significant improvements when using deep structures over shallow structures. However, the differences are not statistically significant for the source and DANN models. 

%All the differences are statistically significant. This is expected as deeper models have higher capacity to model more complex functions. The deep structure offers a relative improvement of 5\% for the within-corpus evaluations (target), 11.7\% for the cross-corpus evaluations (source), and 5.5\% for the proposed DANN framework. When we compare the improvements in CCC for the DANN framework over models trained with the source domain, we observe a relative gain of 30.4\% for the shallow structures and 23.2\% for the deep structures. The deep structures increase the performances of the systems, especially for the models trained with the source domains. Therefore, relative improves are harder to achieve.

\begin{figure}[thb]
    \centering
    \begin{subfigure}[b]{0.4\textwidth}
        \includegraphics[width=0.9\columnwidth]{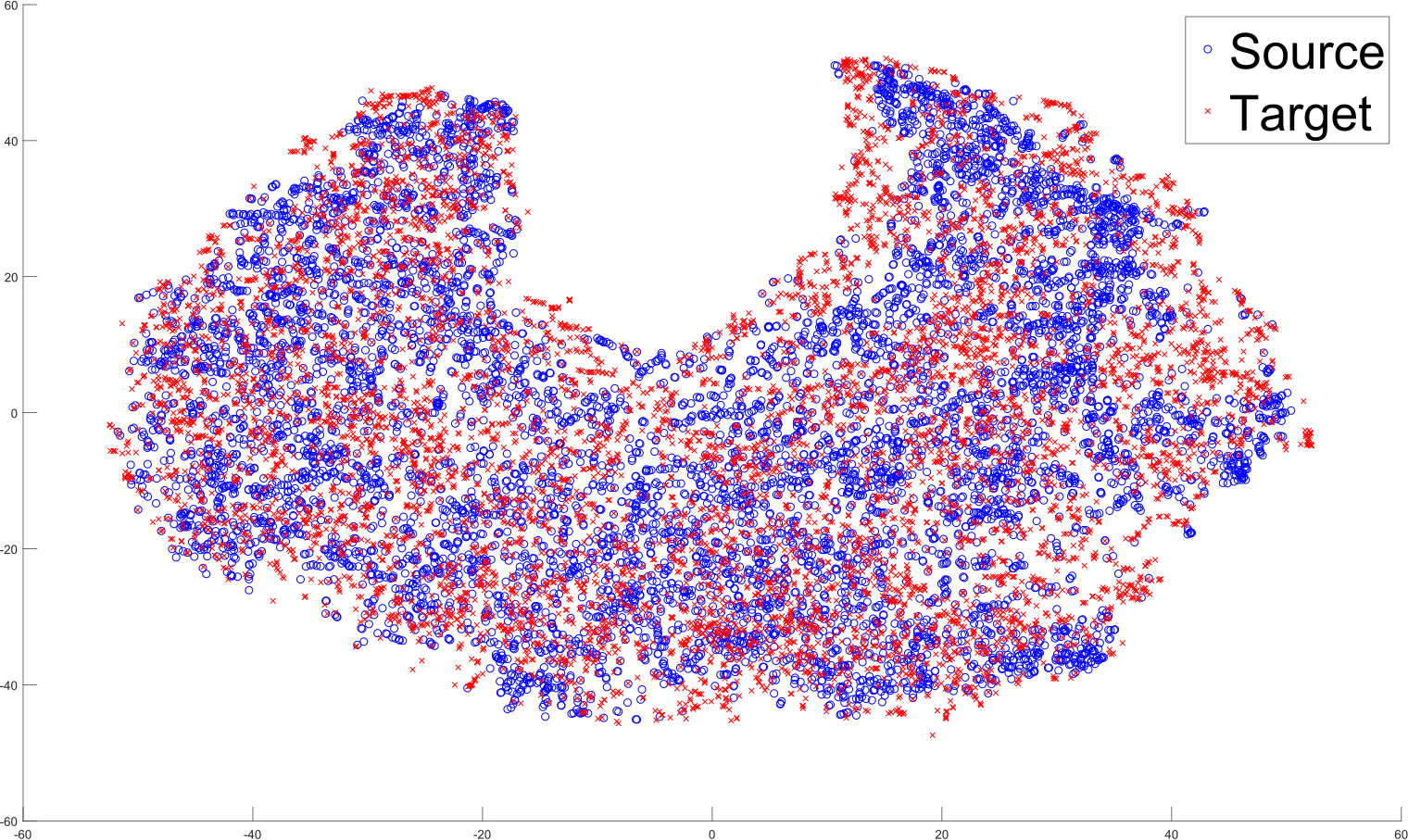}
        \caption{Feature representation using the DANN model}
        \label{fig:dann_rep}
    \end{subfigure}
    
    \begin{subfigure}[b]{0.4\textwidth}
        \includegraphics[width=0.9\columnwidth]{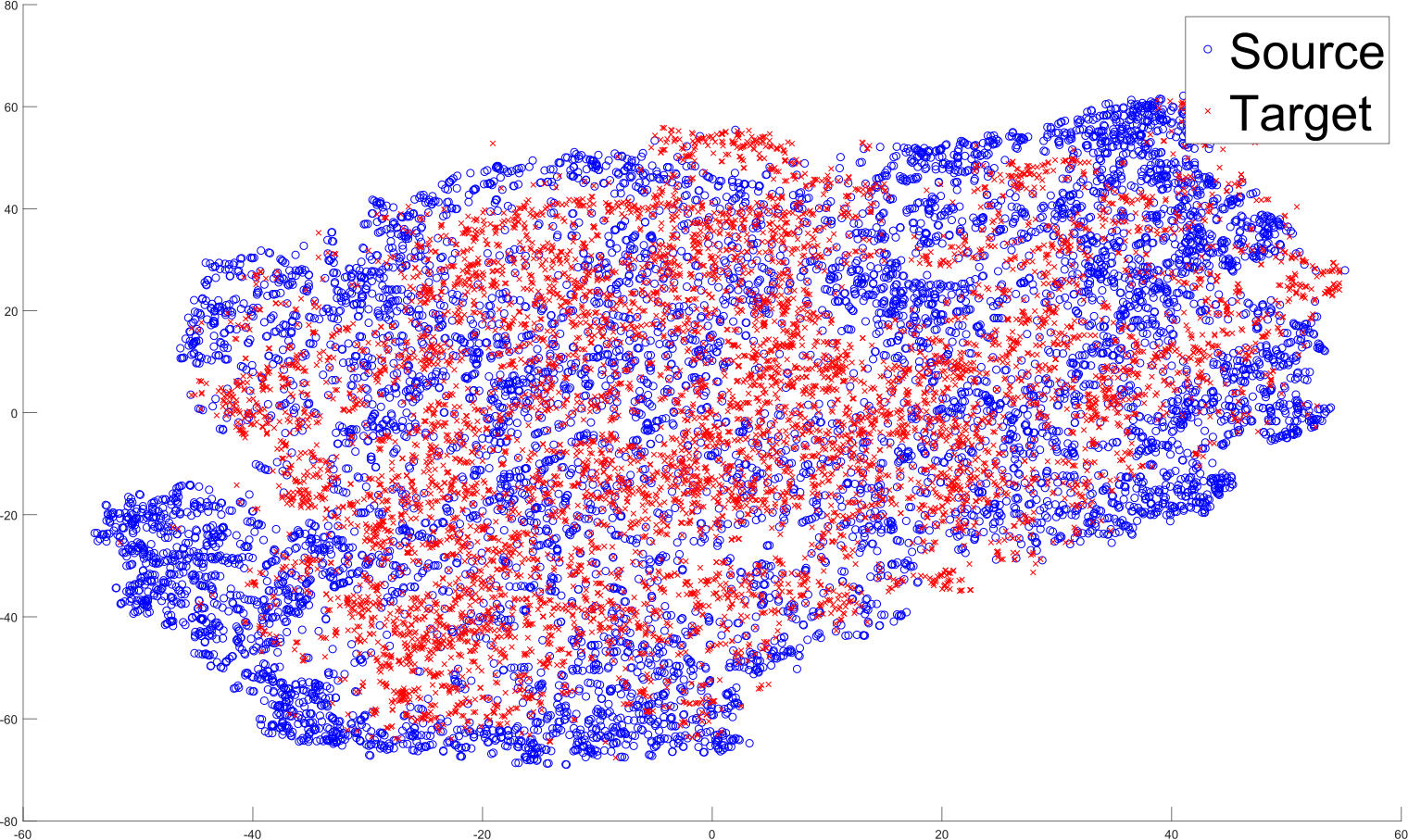}
        \caption{Feature representation using the baseline source model}
        \label{fig:src_rep}
    \end{subfigure}
    \caption{Feature representation at the last shared layer for arousal models trained on USC-IEMOCAP corpus. The figures are created with the t-SNE toolkit.}
    \label{fig:rep}
    \vspace{-0.2cm}
\end{figure}

\begin{figure*}[thb]
    \centering
    \begin{subfigure}[b]{0.23\textwidth}
        \includegraphics[width=\columnwidth]{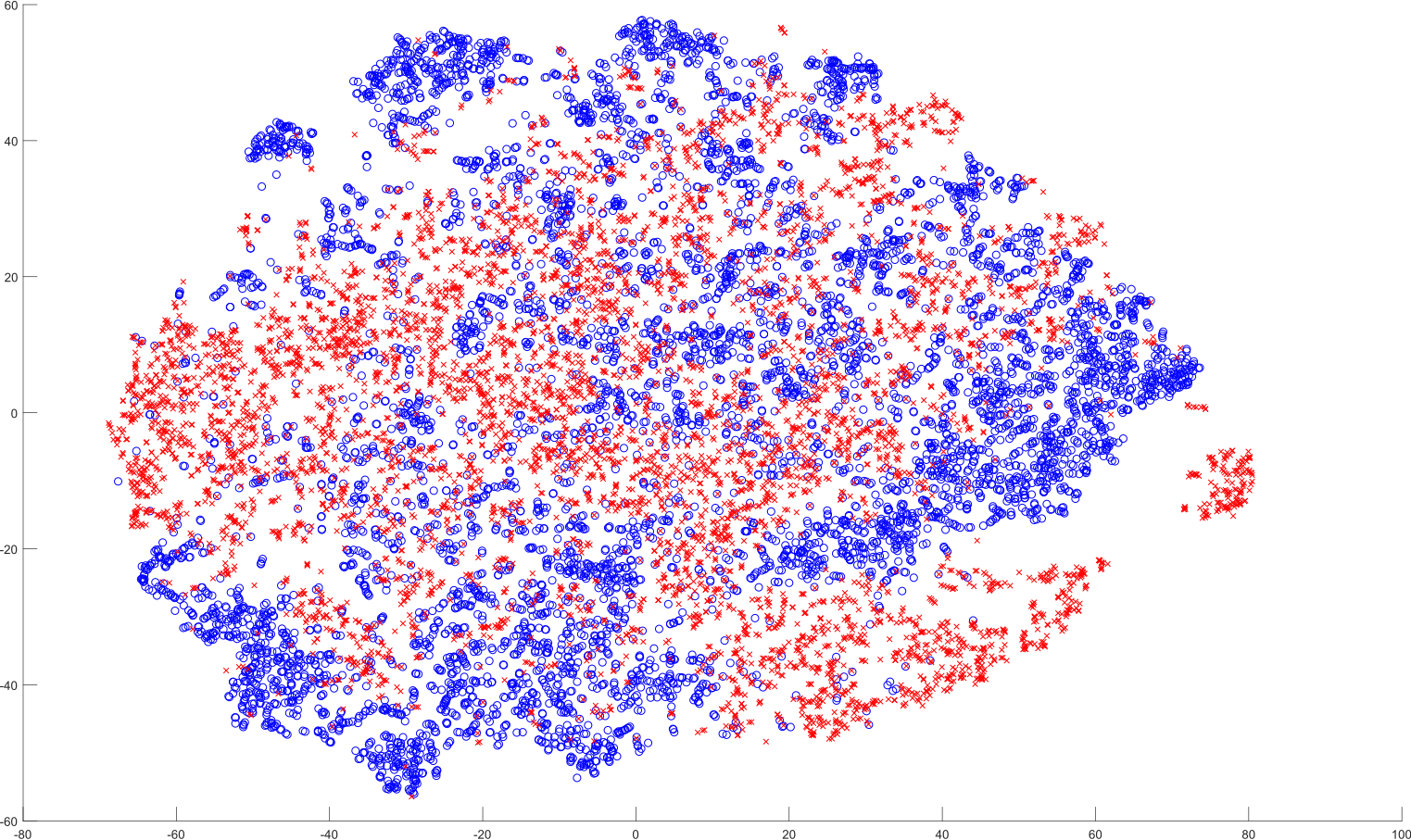}
        \caption{Layer 1 - DANN}
        \label{fig:imp1}
    \end{subfigure}
    ~
    \begin{subfigure}[b]{0.23\textwidth}
        \includegraphics[width=\textwidth]{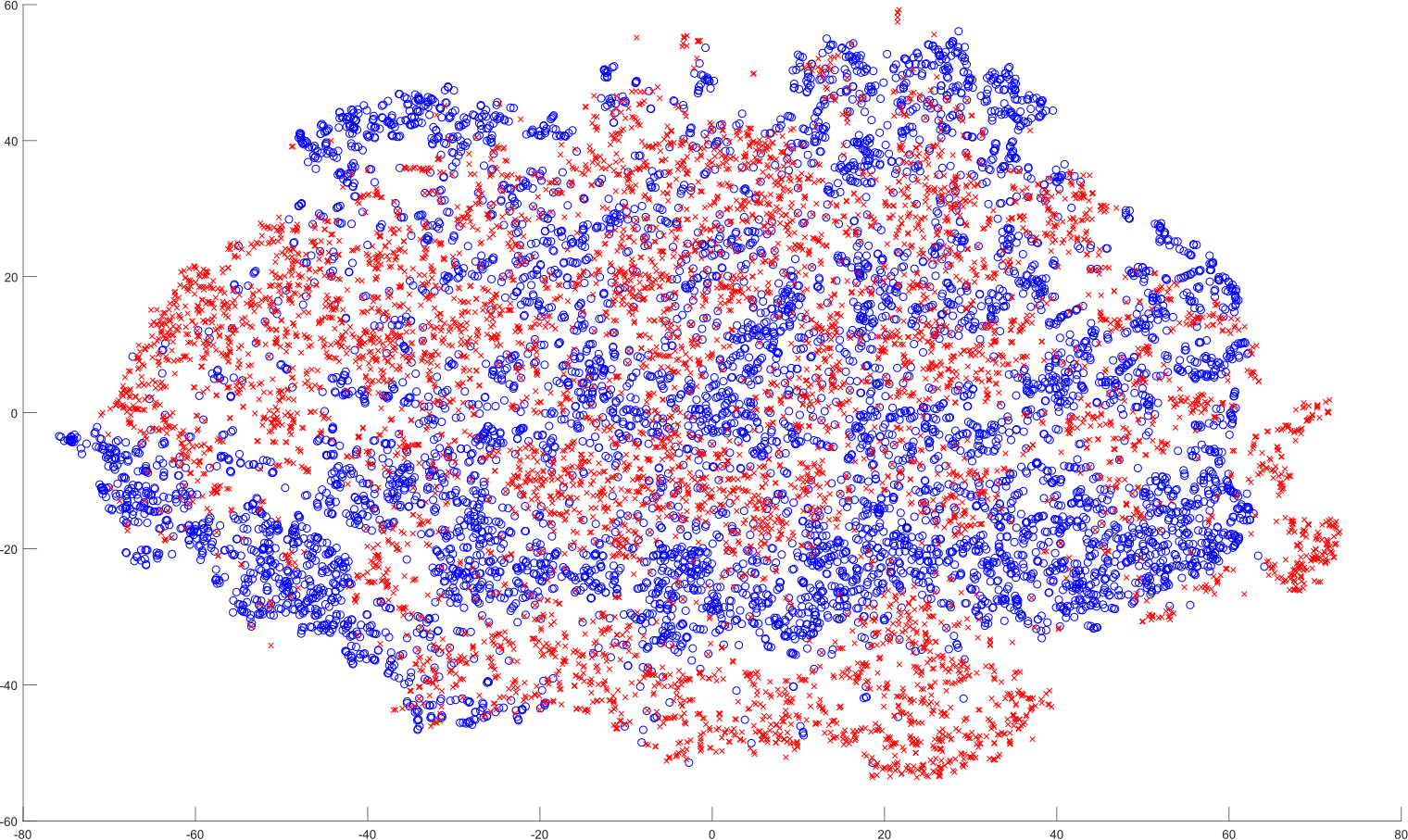}
        \caption{Layer 2 - DANN}
        \label{fig:imp2}
    \end{subfigure}
    ~
    \begin{subfigure}[b]{0.23\textwidth}
        \includegraphics[width=\textwidth]{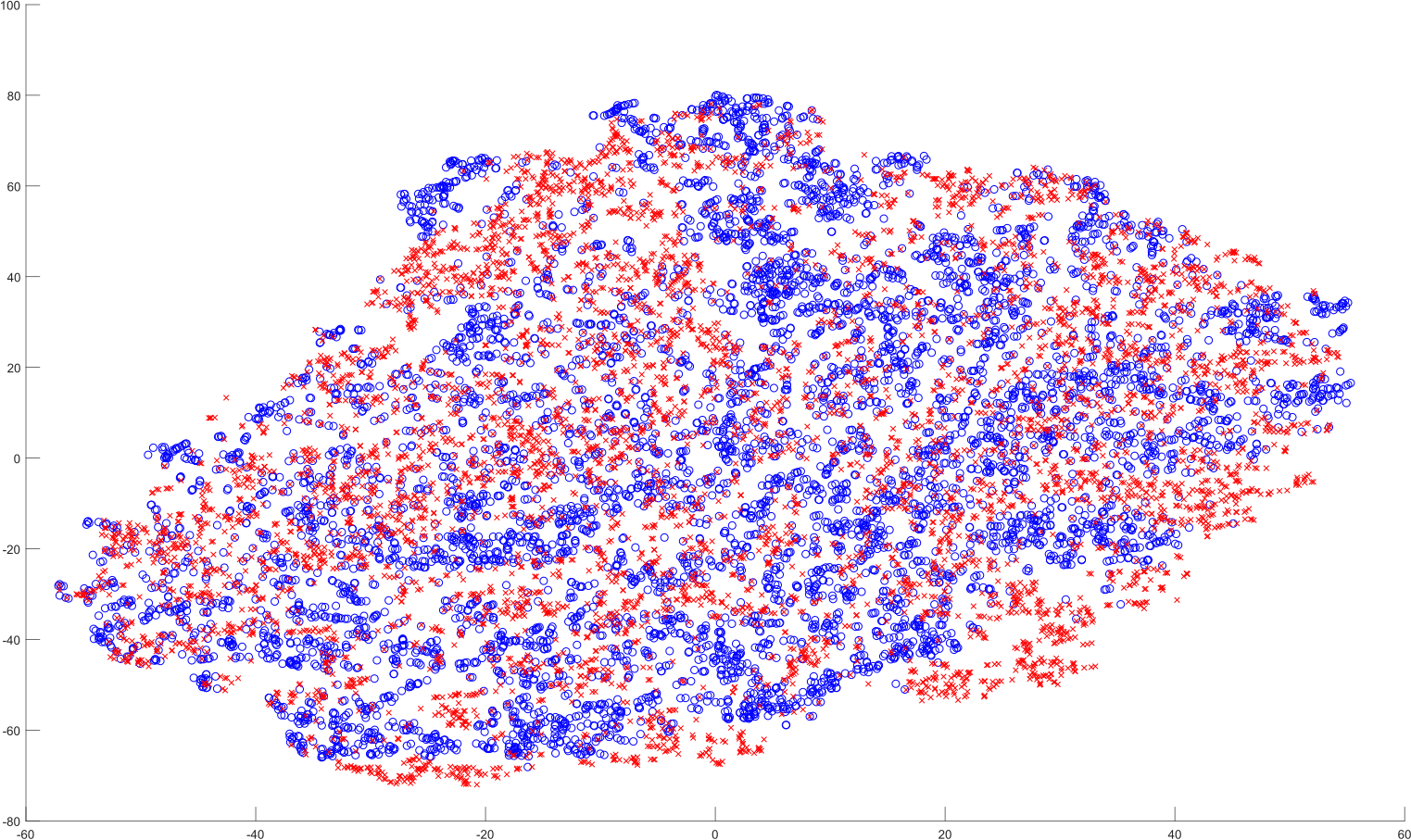}
        \caption{Layer 3 - DANN}
        \label{fig:imp3}
    \end{subfigure}
    ~
    \begin{subfigure}[b]{0.23\textwidth}
        \includegraphics[width=\textwidth]{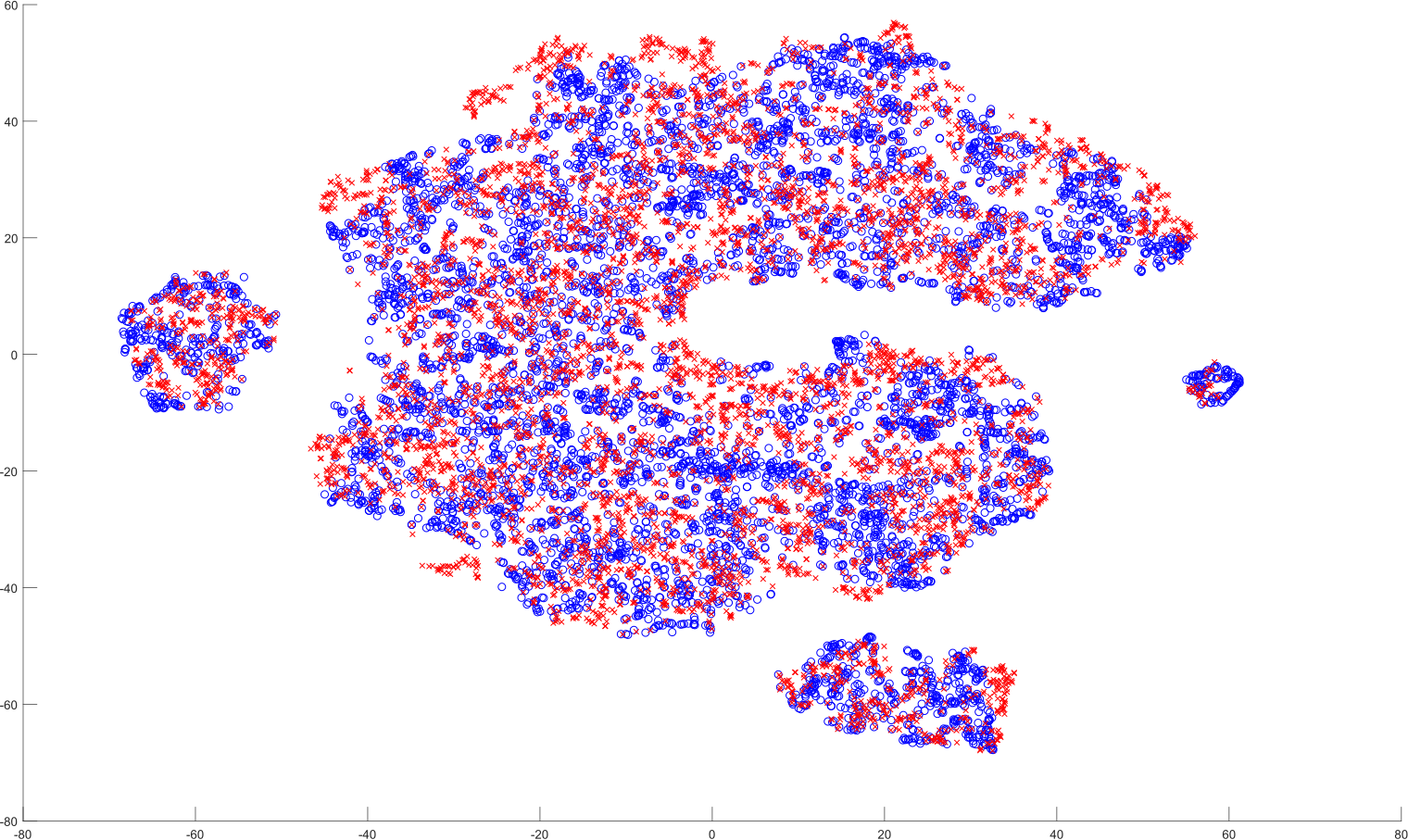}
        \caption{Layer 4 - DANN}
        \label{fig:imp4}
    \end{subfigure}\\
	\begin{subfigure}[b]{0.23\textwidth}
        \includegraphics[width=\columnwidth]{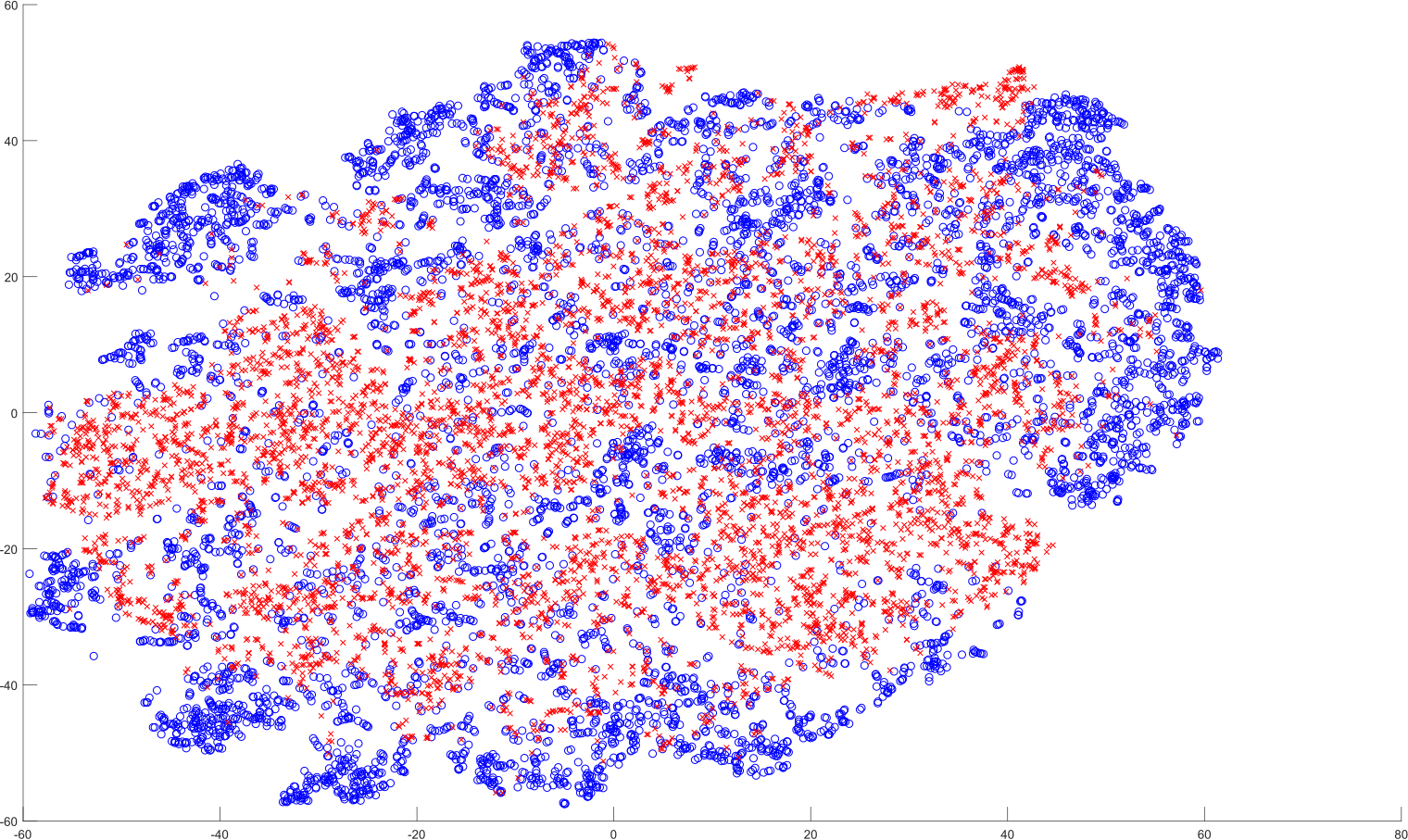}
        \caption{Layer 1 - Source}
        \label{fig:imp1s}
    \end{subfigure}
    ~
    \begin{subfigure}[b]{0.23\textwidth}
        \includegraphics[width=\textwidth]{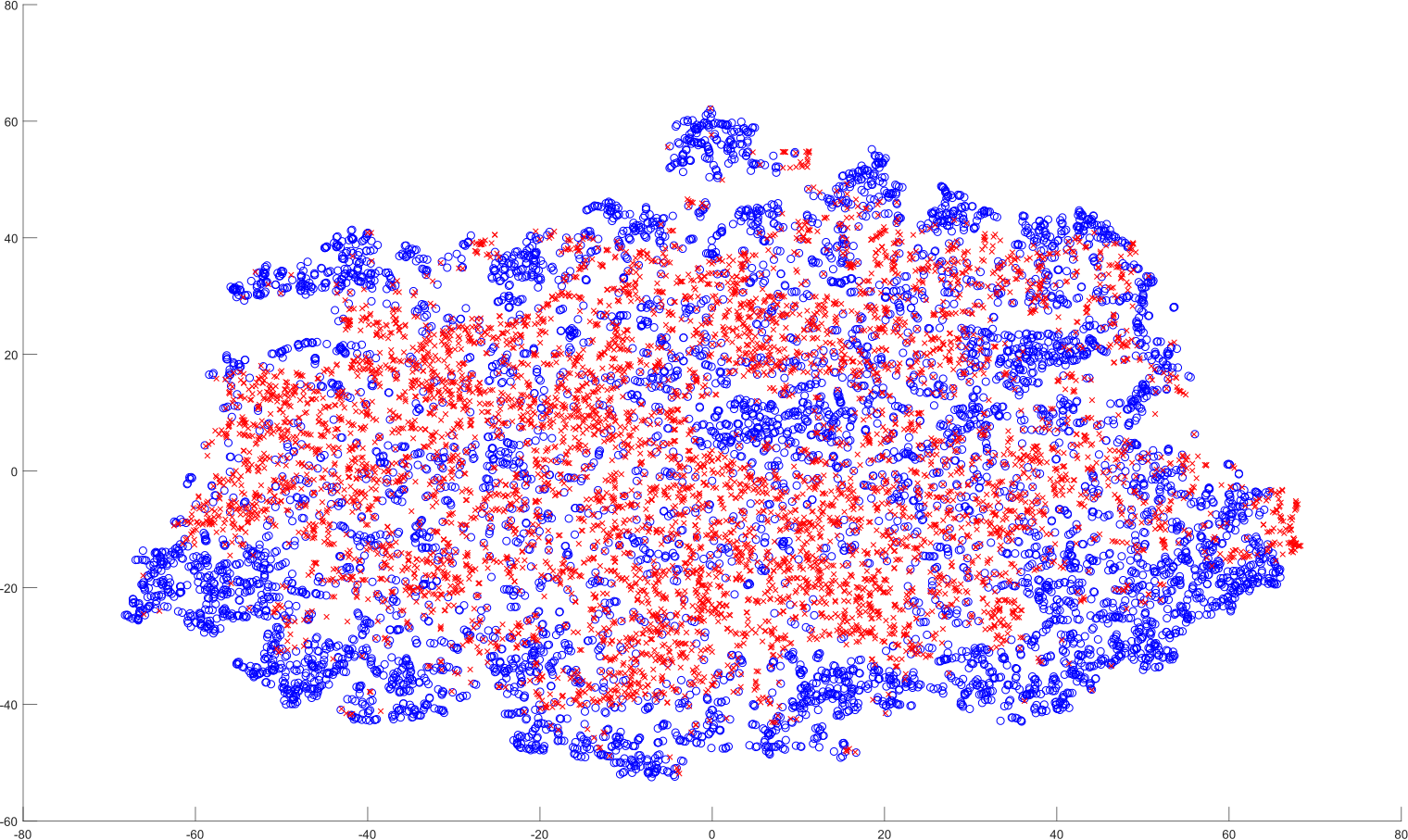}
        \caption{Layer 2 - Source}
        \label{fig:imp2s}
    \end{subfigure}
    ~
    \begin{subfigure}[b]{0.23\textwidth}
        \includegraphics[width=\textwidth]{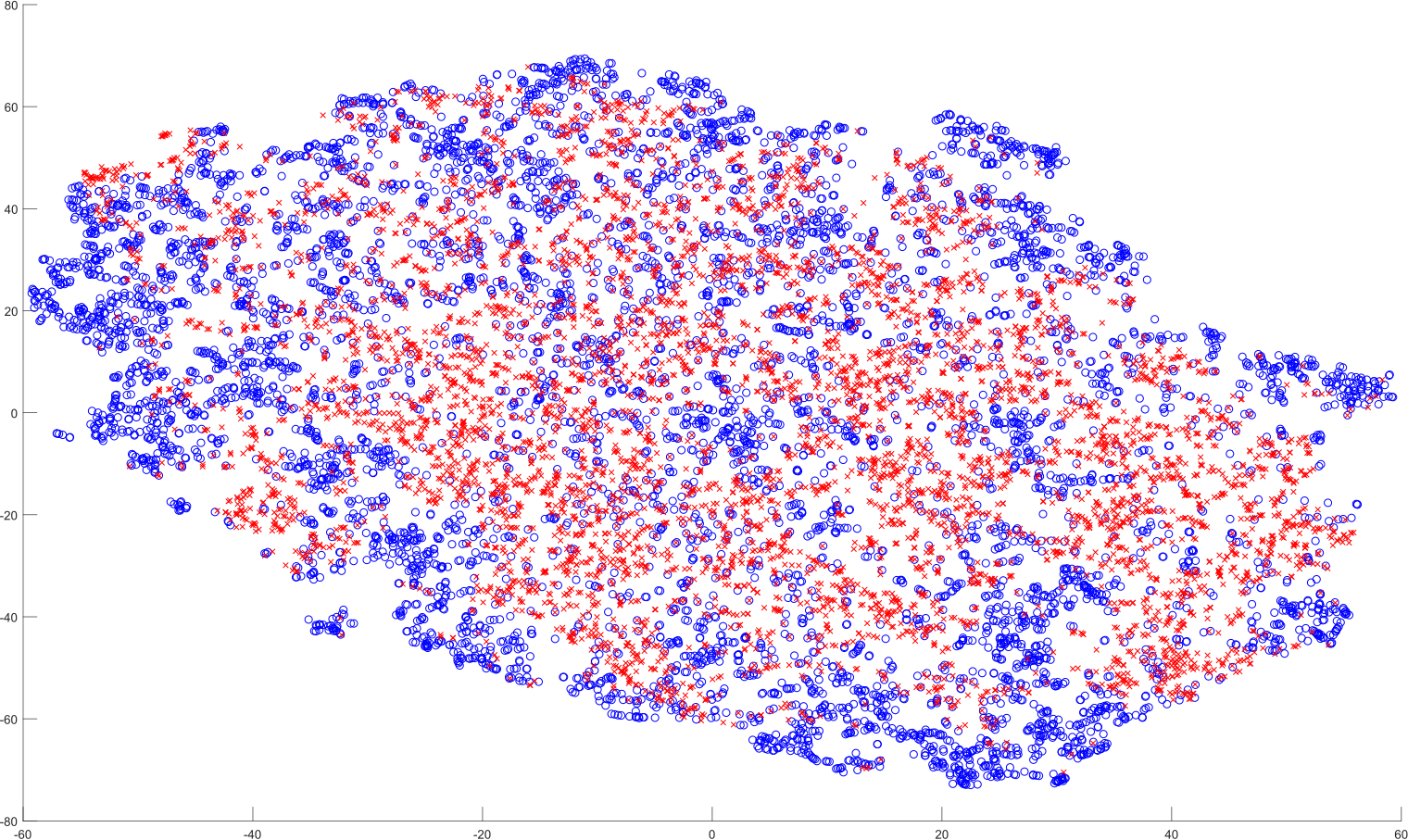}
        \caption{Layer 3 - Source}
        \label{fig:imp3s}
    \end{subfigure}
    ~
    \begin{subfigure}[b]{0.23\textwidth}
        \includegraphics[width=\textwidth]{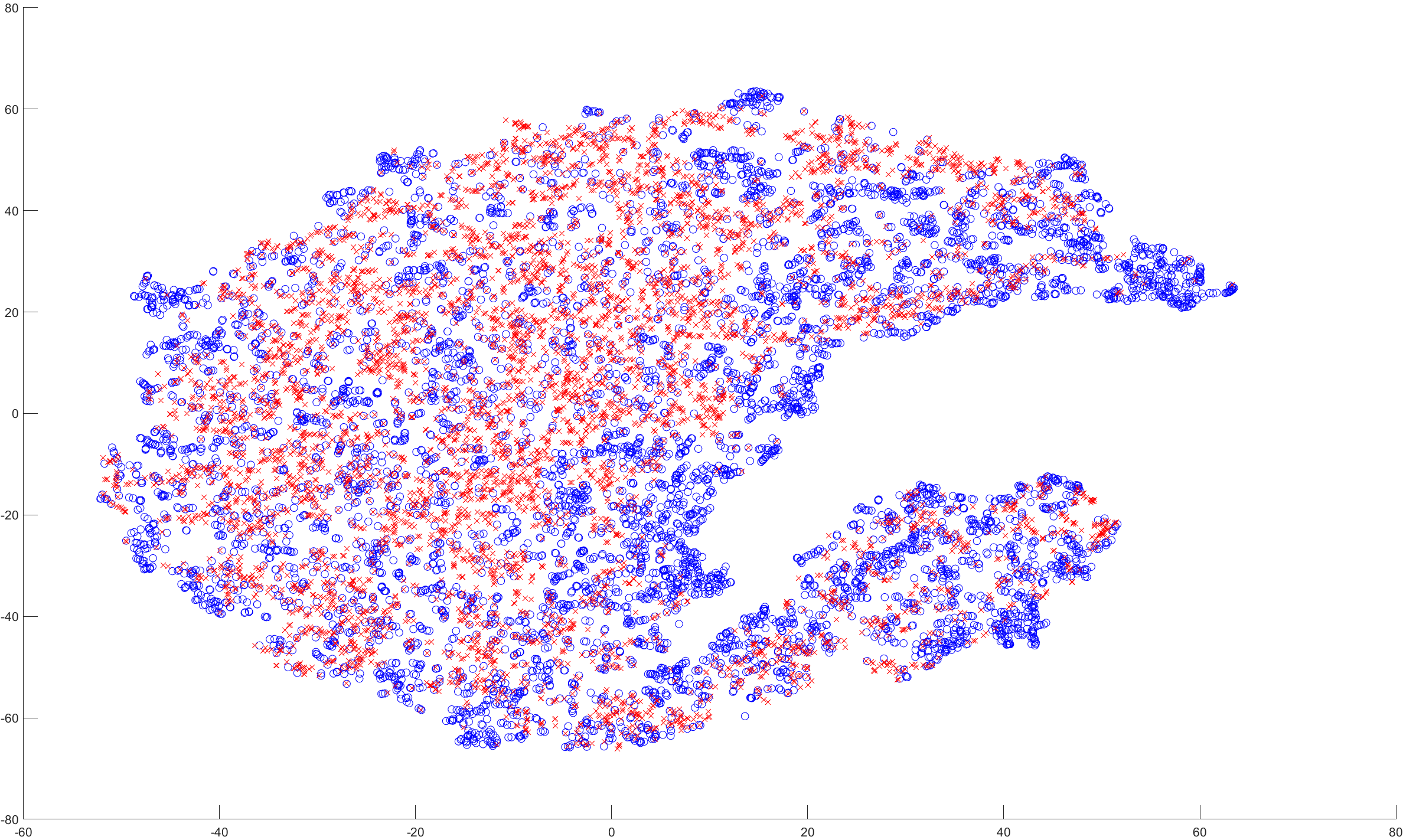}
        \caption{Layer 4 - Source}
        \label{fig:imp4s}
    \end{subfigure}
        \includegraphics[trim=17cm 23cm 16cm 0cm, clip, width=4cm]{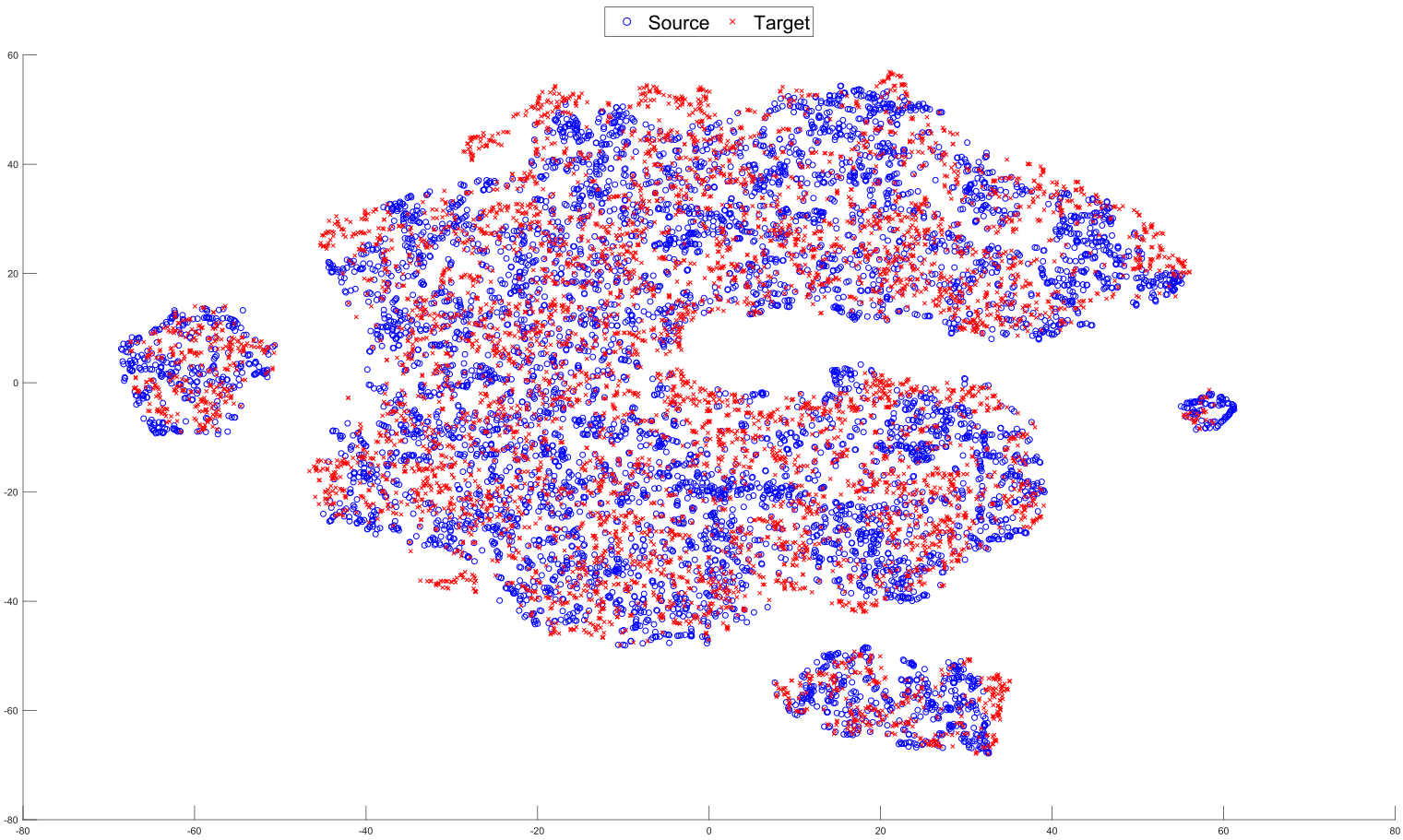}
    \caption{Feature representation at each layer using the DANN model, and the baseline DNN trained with the source domain. The figures are created with the t-SNE toolkit. The example corresponds to models for dominance with four shared layers, using the MSP-IMPROV corpus as the source domain. Figures (a)-(d) report the results for DANN model. Figures (e)-(h) report the results for DNN trained with the source domain.}
    \vspace{-0.2cm}
    \label{fig:rep2}
\end{figure*}

\vspace{-0.2cm}
\subsection{Data Representation}
\label{ssec:datarepresentation}

The results in Tables \ref{tbl:depth} and \ref{tab:1} demonstrate the benefits of using the proposed DANN framework. This section aims to understand the key aspect of the DANN approach by visualizing the feature representation created during the training process.

We use the t-SNE toolkit \cite{Maaten_2014} to create 2D projections of the feature distributions at different layers of the networks. Figure \ref{fig:rep} shows the distributions of the data from the source domain (blue circles) and the target domain (red crosses) after projecting them in the feature representation created by two models. This example corresponds to the models trained for arousal using the USC-IEMOCAP corpus as the source domain (as explained in Section \ref{ssec:depth}, the DANN model for arousal has only one shared layer as a feature representation). Figure \ref{fig:dann_rep} shows the data representation at the shared layer of the DANN model. Figure \ref{fig:src_rep} shows the data representation at the equivalent layer of the DNN model trained with the source domain (i.e., the USC-IEMOCAP database). By using adversarial domain training, the feature distribution for samples from both domains are almost indistinguishable, demonstrating that the proposed approach is able to find a common representation. Without adversarial domain training, in contrast, there are large regions in the feature space where it is easy to separate samples from the source and target domains. Figure \ref{fig:src_rep} suggests the presence of a source-target mismatch which affects the performance of the emotion classifier.

We also explore the feature representation when the DANN model is trained with four shared layers using the t-SNE toolkit. The objective of this evaluation is to visualize the distribution of the data in each of the shared layers. This evaluation is implemented using the models for dominance, using the MSP-IMPROV corpus as the source domain. Figures \ref{fig:imp1}-\ref{fig:imp4} show the changes in the data representation for each of the four shared layers in the DANN model. At the first layer (Fig. \ref{fig:imp1}), the feature representation for the source data (blue circles) and the target data (red crosses) are dissimilar enough for the domain classifier to be able to distinguish them.  While the difference between domains in the feature representation has decreased at the second layer (Fig. \ref{fig:imp2}), there are still some regions dominated by samples from one of the two domains. At the third layer (Fig. \ref{fig:imp3}), the feature representation of the domains are similar enough to confuse the domain classifier. The common representation is maintained in the fourth layer (Fig. \ref{fig:imp4}), where the data from the target domain is indistinguishable from the data from the source domain. This final representation is used by the emotion regression to predict the emotional state of the target speech. For comparison, we also trained a baseline model with six layers, matching the combined number of shared and task classifier layers in the DANN model. Figures \ref{fig:imp1s}-\ref{fig:imp4s} show the feature representation of the corresponding first four layers of this model. In the DANN model, the data representation of the samples from the source and target domains become more similar in deeper layers. This trend is not observed in the model trained with only the source data. The DANN model effectively reduces the mismatch in the feature representation across domains, which leads to significant improvements in the regression models.

%\pagebreak
\vspace{-0.2cm}
\section{Conclusions}
\label{sec:conc}

This study proposed an appealing framework for emotion recognition that exploits available unlabeled data from the target domain. The proposed approach relies on \emph{domain adversarial neural network} (DANN), which creates a flexible and discriminant feature representation that reduces the gap in the feature space between the source and target domains. By using adversarial training, we learned domain invariant representations that can effectively discriminate the primary regression task. The model aims to find a balanced representation that aligns the domain distributions, while retaining crucial information for the  primary regression task. The proposed adversarial training leads to significant improvements in the emotion recognition classifier's performance over models exclusively trained with data from the source domain, which was demonstrated by the experimental evaluation. We visualized the data representation of both domains by projecting the features into the shared layers of the proposed DANN model. The results showed that the model converged to a common representation, where the source and target domains became indistinguishable. The experimental evaluation also showed that the amount of labeled data from the source domain plays a small role in determining how many shared layers are needed between the domain and regression tasks. Since the number of shared layers has a strong impact on the system's performance, it is vital to identify the optimal number of shared layers, given a specific source domain.

One challenging aspect in using the proposed approach is the difficulty of training adversarial networks. For example, Shinohara \cite{shinohara_2016} noted that in ASR problems, the improvements of DANN were large for some types of noises, but less effective for others. They suggested that tuning the parameters could lead to better results. We also observed that the framework failed to converge for certain parameters, which is common in minimax problems. When properly trained, however, this powerful framework can elegantly solve one of the most important problems in speech emotion recognition: reducing the mismatch between train and test domains.

In the case of multiple sources, our approach seems to work well when multiple sources are combined, treating them as one. This approach forces the network to learn a representation that is common across all the source domains. We hypothesize that a better approach is to use asymmetric transformations, where the model learns multiple possible representations for the test data, creating one representation for each source. During testing, the network chooses the most useful representation for each data point. Another alternative approach is to transform the available sources to match the target domains. Finally, this unsupervised approach can be easily extended to the cases where limited labeled data from the target domain is available (semi-supervised approach), creating a flexible framework to create emotion recognition systems that can generalize across domain.

% or


%\appendices
%% you can choose not to have a title for an appendix
%% if you want by leaving the argument blank
%\section{}

\vspace{-0.2cm}
\section*{Acknowledgment}

This study was funded by the National Science Foundation (NSF) CAREER grant IIS-1453781.
\vspace{-0.3cm}

\ifCLASSOPTIONcaptionsoff
  \newpage
\fi



%\bibliographystyle{IEEEtran}
%\begin{thebibliography}{1}
%\end{thebibliography}

\bibliographystyle{IEEEbib}
\bibliography{reference}

\end{document}